\documentclass[prd,twocolumn,amsmath,amssymb,floatfix, superscriptaddress]{revtex4}
\usepackage{graphicx, epsfig, bm,amsmath,natbib}

\usepackage{color}
\usepackage{hyperref}
\usepackage{ifthen}
\usepackage{xstring}
\usepackage{graphicx,epsfig}
\usepackage{subfigure}
\usepackage{amssymb}
\usepackage{afterpage}

\def\be{\begin{equation}}
\def\ee{\end{equation}}
\def\ba{\begin{eqnarray}}
\def\ea{\end{eqnarray}}

\newcommand{\Imax}{I_{1,{\rm max}}}

\begin{document}

\title{Complete WMAP Constraints on Bandlimited Inflationary Features}

\author{Cora Dvorkin}
\affiliation{Kavli Institute for Cosmological Physics, Enrico Fermi Institute,
        University of Chicago, Chicago, IL 60637}
\affiliation{Department of Physics, University of Chicago, Chicago, IL 60637}

\author{Wayne Hu}
\affiliation{Kavli Institute for Cosmological Physics, Enrico Fermi Institute,
        University of Chicago, Chicago, IL 60637}
\affiliation{Department of Astronomy \& Astrophysics, University of Chicago, Chicago, IL 60637}

\begin{abstract}
  Using a principal component (PC) basis  that accommodates  order unity
features in the slow roll parameters as fine as 1/10 of a decade  across more than 
2 decades of the inflationary expansion, we test slow roll and single field inflation 
with the WMAP7 data.   Detection of any non-zero component would represent a 
violation of ordinary slow roll and indicate a feature in the inflaton potential or sound speed.   Although one component
shows a deviation at the 98\% CL, it cannot be considered statistically significant given the
20 components tested.   
The maximum likelihood PC parameters only improves $2\Delta \ln L$ by 17 for the 20 
parameters associated with known glitches in the WMAP power spectrum at multipoles $\ell < 60$.
We make model-independent predictions for the matching glitches in the polarization spectrum that would test their
inflationary origin.
This complete analysis for bandlimited features in the source function of generalized slow roll
can be used to constrain parameters of specific models of the
inflaton potential without requiring a separate likelihood analysis for each choice.  We illustrate its use by placing bounds on the height and width of a step-like feature in the potential proposed to explain the glitch at $20 \le \ell \le 40$.  
Even allowing for the presence of features in the temperature spectrum, single field inflation makes sharp falsifiable predictions for the acoustic peaks in
the polarization whose violation would require extra degrees of freedom.
\end{abstract}
\maketitle

\section{Introduction}

Observed glitches in the WMAP power spectrum of cosmic microwave background (CMB) temperature
fluctuations  \cite{Peiris:2003ff} have motivated many studies of features in the initial conditions.
Most of these studies have focused on the reconstruction of the curvature power spectrum 
through parametric, minimally parametric or regularized inverse techniques 
(e.g.~\cite{Hu:2003vp,Bridle:2003sa,Hannestad:2003zs,Leach:2005av,TocchiniValentini:2004ht,
Mukherjee:2005dc,Bridges:2006zm,Shafieloo:2007tk,Peiris:2009wp,Nicholson:2009zj,Nagata:2008zj}).

Reconstruction of the curvature spectrum suffers from two potential issues.   Given that
fine scale features are observable at high wavenumber,  parametric
models are not complete unless a very large number of parameters are employed.
Secondly, not all curvature power spectra can arise from physical mechanisms in
the early universe making parametric models potentially overcomplete and subject
to fitting the noise instead of fitting the physics.   For example, a delta function in
the initial curvature spectrum would be highly observable but not expected to arise
in any physical model. 

For the purposes of testing inflationary models of the initial conditions  one can instead
try to constrain the shape of the inflaton potential under the assumption that inflation arises from
a single scalar field with a canonical kinetic term.  Specific potentials have been used to test the origin of the low quadrupole moment,
the glitches at multipole moments $\ell=20-40$  and glitches near the WMAP beam scale
\cite{Peiris:2003ff,Contaldi:2003zv,Covi:2006ci,Hamann:2007pa}.  On the other hand, 
model-independent reconstruction 
approaches have implicitly or explicitly assumed a slowly varying inflaton potential
 \cite{Lidsey:1995np,Grivell:1999wc,Leach:2003us,Peiris:2006sj,Kinney:2008wy}.

Sharp features in the inflaton potential cause features in the temperature power
spectrum \cite{Starobinsky:1992ts,Adams:2001vc}.   As long as those features are of small amplitude,
 inflation continues uninterrupted but certain slow roll parameters are neither constant
nor necessarily small.   The generalized slow roll (GSR) approach \cite{Stewart:2001cd,Choe:2004zg,Kadota:2005hv,Dvorkin:2009ne} can be used to analyze such cases.   In particular to good approximation there is a single source function that encodes observable features
in the inflation potential \cite{Dvorkin:2009ne} for canonical kinetic terms or the
sound speed for non-canonical terms \cite{Hu11}. This function is also closely related to the source of corresponding bispectrum features \cite{Adshead:2011bw}.  In previous work, we studied the strong constraints on this function imposed by the precise and featureless measurements
around the first acoustic peak through a low order principal components decomposition
\cite{DvoHu10a}. 

In this paper, we extend our previous analysis to a  basis of 
20 principal components for the source function of inflationary features.    This basis is complete for models where the features vary no more rapidly than 10 per decade of the
expansion or about 4 per efold during inflation.
In \S \ref{sec:GSR} we review the GSR and principal components technique.
  In Appendix \ref{sec:MCMCopt}, we describe numerical techniques used to reduce the computation time of the analysis.  We test the validity of the GSR approximation in
  Appendix \ref{sec:GSRaccuracy}.
We present the results of the WMAP likelihood analysis in \S \ref{sec:results}.
In \S \ref{sec:apps} we develop tests of single field inflation and consider applications
to specific classes of potentials.    We discuss these results in \S \ref{sec:discussion}.

\section{Methodology}
\label{sec:GSR}

We use the generalized
slow roll (GSR) approximation and principal components to study features generated by single-field inflation.  
We  refer the reader to Ref.~\cite{DvoHu10a} for details but provide
a brief description here.
Features arise from a single source function 
that describes the deviation from slow roll in the background 
(\S \ref{sec:GSRbasics}) which we can be decomposed into a 
 basis of principal components that is complete for bandlimited models sampled at a maximal rate per efold (\S \ref{sec:GSRpc}).

\subsection{Generalized Slow Roll}
\label{sec:GSRbasics}

Under the GSR approximation, features in the curvature power spectrum are generated
by a single source function of the background evolution of the inflaton $\phi$
\ba
G'(\ln\eta)&\equiv& -2 (\ln f)' + {2 \over 3} (\ln f)'' \,,
\label{eqn:sourcefunction}
\ea
where 
\begin{equation}
f = {2\pi \dot \phi a \eta \over H}\,.
\end{equation}  
 Primes here and below denote derivatives with respect to $\ln \eta$ where 
$\eta=\int_t^{t_{\rm end}} dt'/a(t')$ 
is the conformal time to the end of inflation and we take units of $M_{\rm pl} = (8\pi G)^{-1/2}=1$.

In the ordinary slow roll approximation, the curvature
power spectrum is given by $\Delta_{\cal R}^2 \approx f^{-2}$ since
\begin{equation}
\epsilon_H = {1\over 2} \left({ \dot \phi \over H} \right)^2, \quad \eta \approx {1 \over aH}.
\end{equation}
 In the GSR approximation,  the curvature power spectrum is instead determined by
 features in the source function through
\begin{eqnarray}
\label{eqn:ourGSR}
\ln \Delta_{\cal R}^2(k) &  \approx  & G(\ln \eta_{\rm min}) +  \int_{\eta_{\rm min}}^{\eta_{\rm max}} {d \eta \over \eta}  W(k\eta) G'(\ln \eta) \nonumber\\
&&+ \ln [ 1 + I_1^2(k) ]\,,
\end{eqnarray}
where integrating $G'$ gives
\begin{equation}
G(\ln\eta)=-2 \ln f + {2\over 3} (\ln f)'.
\end{equation}
Here the non-linear correction is given by 
 \begin{eqnarray}
 I_1(k) =
{1 \over \sqrt{2}}   \int_{\eta_{\rm min}}^{\eta_{\rm max}} {d \eta \over \eta} X(k\eta) G'(\ln \eta)   \,. 
\end{eqnarray}
We take $\eta_{\rm min}=1$ Mpc  and $\eta_{\rm max} = 10^5$ Mpc which more than covers
the range observable to WMAP.
The window functions 
\begin{eqnarray}
W(u) &=& {3 \sin(2 u) \over 2 u^3} - {3 \cos (2 u) \over u^2} - {3 \sin(2 u)\over 2 u} \,, \nonumber\\
X(u) &=& {3 \over u^3} (\sin u - u \cos u)^2 \,,
\end{eqnarray}
define the linear and nonlinear response of the curvature spectrum to $G'$ respectively.  
Accuracy of the GSR approximation requires the nonlinear response, as quantified by $I_1^2$, to remain below order unity. We call this the GSR condition (see Appendix \ref{sec:GSRaccuracy}).
\begin{figure}[tbp]
\includegraphics[width=0.45\textwidth]{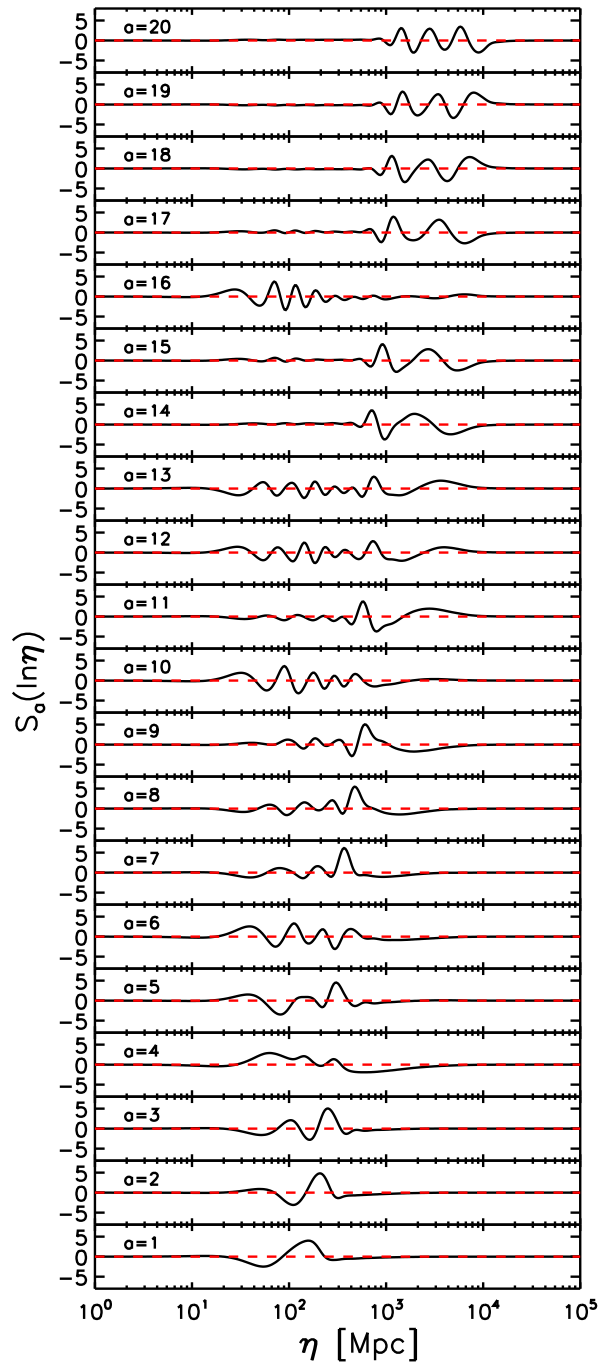}
\caption{The first 20 principal components of the GSR source $G'$ as a function of conformal time to the end of inflation, in order of increasing variance from bottom to top. 20 PC components suffice to represent inflationary
features observable to WMAP that vary no more rapidly than $\sim 1/4$ of an efold.  Here and below, dashed red lines represent power law conditions with zero 
amplitude in the PC components.}
\label{plot:20PCs}
\end{figure}

\subsection{Principal Component Analysis}
\label{sec:GSRpc}

The principal components of the WMAP7 Fisher matrix provide an efficient
basis with which to decompose the source function (\ref{eqn:sourcefunction})
\begin{equation}
G'(\ln \eta)=1- n_s + \sum_{a=1}^N m_a S_a(\ln \eta) \,,
\label{eq:Gprime_reconstruction}
\end{equation}
where the eigenfunctions $S_a$ are constructed following Ref.~\cite{DvoHu10a} by sampling at a rate of 10 per decade in $\eta$ or equivalently 4.3 per efold of inflation
across $\eta = [1-10^5]$Mpc.  
In terms of the width of features in the potential, this limit corresponds to $\Delta \phi 
\gtrsim \epsilon_H^{1/2}/4.3.$
 This rate is sufficient to capture models that describe the glitches in the WMAP7 power spectrum
(see \S \ref{sec:apps} for a discussion of the limitations imposed by the sampling). 

The amplitudes $m_a$ then can be incorporated into a  Markov Chain Monte Carlo (MCMC)
likelihood analysis
of the WMAP data.
As described in Appendix \ref{sec:MCMCopt} we slightly modify the original
approach \cite{DvoHu10a} to improve the convergence properties of the MCMC
analysis.
  Since a constant $G'$ described by 
$n_s$ is equivalent
to tilt in the curvature spectrum and $G(\ln\eta_{\rm min})$ is equivalent to a normalization
parameter we replace them with effective parameters $\bar G'$ and $A_c$.  
Specifically $\bar G'$ is an average of $G'$ for $30 < \eta/{\rm Mpc}< 400$ and $A_c$ is the normalization of the temperature power spectrum $C_\ell^{TT}$ at the first peak $\ell=220$ relative to a
fiducial choice that fits the WMAP7 data.
From these two phenomenological parameters
we can derive constraints for the tilt $n_s$ and curvature power spectrum normalization $A_s$ (see Appendix \ref{sec:MCMCopt}).

Since a signal-to-noise analysis shows that $20$ out of the 50 principal components are required for a complete representation of the WMAP data at our bandlimit \cite{DvoHu10a}, we choose $N=20$ for our analysis.
These first $20$ principal components are shown in Fig. \ref{plot:20PCs}. 
Note that the first 10 components resemble local Fourier modes around $\eta \approx
10^2$ Mpc where the well-constrained first acoustic peak gets its power.
It is not until components 11-20 that horizon scale features at low multipole or $10^3-10^4$ Mpc are
represented.

We use the MCMC method 
to  determine joint constraints on the 20 PC amplitudes and 
cosmological parameters
\begin{equation}
p_{\mu}=\{m_1,\ldots,m_{20},A_{c},\bar G,\tau,\Omega_bh^2,\Omega_{c}h^2,\theta\}\,.
\label{eq:parameters}
\end{equation}
Here $\tau$ is the reionization optical depth, $\Omega_b h^2$ is the physical
baryon density, $\Omega_c h^2$ is the physical dark matter density and $\theta$ is
the angular size of the sound horizon at recombination.

  The MCMC algorithm samples
the parameter space  evaluating the
likelihood ${\cal L}({\bf x}|{\bf p})$ of the data  ${\bf x}$  given each proposed parameter set ${\bf p}$
({\it e.g.}\ see~\cite{Christensen:2001gj,Kosowsky:2002zt}).  
The posterior distribution is obtained using Bayes' Theorem,
\begin{equation}
{\cal P}({\bf p}|{\bf x})=
\frac{{\cal L}({\bf x}|{\bf p}){\cal P}({\bf p})}{\int d\bm{\theta}~
{\cal L}({\bf x}|{\bf p}){\cal P}({\bf p})},
\label{eq:bayes}
\end{equation}
where ${\cal P}({\bf p})$ is the
prior probability density.  We place non-informative tophat priors on all parameters in
Eq.~(\ref{eq:parameters}).  To ensure the validity of the GSR approximation we
also place a tophat prior on
\begin{equation}
I_{1,{\rm max}}={\rm max}|I_1(k)|.
\label{eq:I1max}
\end{equation}
As shown in Appendix \ref{sec:GSRaccuracy}, a value of $I_{1,{\rm max}}=1/\sqrt{2}$ is sufficient
to ensure accuracy of the GSR approximation.

\begin{figure}[tbp]
\includegraphics[width=0.475\textwidth]{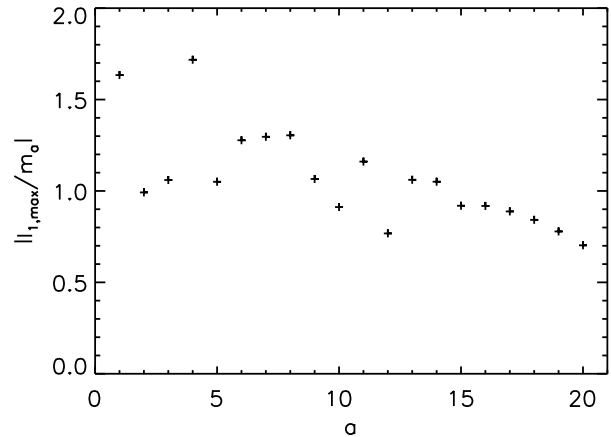}
\caption{Sensitivity of  the nonlinearity parameter 
$\Imax$ (see Eq.~\ref{eq:I1max}) to the amplitude of the first 20 PCs considered individually.
This parameter must be less than order unity for the GSR approximation to be
accurate, and we typically place a prior of $\Imax<1/\sqrt{2}$.}
\label{plot:I1max_over_ma}
\end{figure}

Fig.~\ref{plot:I1max_over_ma} shows the maximal contribution to $I_1$ per unit
amplitude deviation in
each of the first 20 principal components.   The higher PCs actually
produce a slightly smaller response largely because the frequency of the oscillations
in Fig.~\ref{plot:20PCs} begins to exceed that of the nonlinear response function $X(k\eta)$.  Thus a prior of $I_{1,{\rm max}}=1/\sqrt{2}$ actually allows high PC components to reach
order unity and $|G'|$ to reach $\sim 4$ or greater.

The MCMC algorithm generates random draws from the
posterior distribution that are fair samples of the likelihood surface.
We test convergence of the samples to a stationary distribution that
approximates the joint posterior density ${\cal P}({\bf p}|{\bf x})$ 
by applying a conservative Gelman-Rubin criterion \cite{gelman/rubin}
of $R-1< 0.01$ across four chains.
We use the
code CosmoMC \cite{Lewis:2002ah} for the MCMC analysis
\footnote{\url{http://cosmologist.info/cosmomc}}.

\begin{table*}[tbp]
\centering
\begin{tabular}{|l|r@{}|c|c|r@{}|c|c|}
\hline
Parameters & & \multicolumn{2}{|c|}{All Data} & & \multicolumn{2}{|c|}{CMB Only}\\
\cline{1-1} \cline{3-4} \cline{6-7}
$100\Omega_b h^2$  & & $2.241\pm0.048$ & $2.233$ 
& & $2.231\pm0.051$ & $2.229$
\\
$\Omega_c h^2$     & & $0.1101\pm0.0040$ & $0.1098$ 
 & & $0.1110\pm0.0051$ & $0.1116$
\\
$\theta$           & & $1.0398\pm0.0022$ & $1.0397$ 
& & $1.0394\pm0.0022$ & $1.0394$
\\
$\tau$             & & $0.089\pm0.014$ & $0.086$ 
& & $0.087\pm0.015$ &  $0.085$
\\
$n_s, 1-\bar G'$              & & $0.9669\pm0.9882$ & $0.9649$  
& & $0.9620\pm0.0078$ & $0.9622$ 
\\
$\ln[10^{10}A_s]$  & & $3.0808\pm0.0332$ & $3.0733$ 
&& $3.0770\pm0.0338$ & $3.0746$
\\
\cline{3-4} \cline{6-7}
$H_0$              & & $71.23\pm1.74$ & $71.23$
& & $70.70\pm2.32$ & $70.40$
 \\
$\Omega_\Lambda$ && $0.738\pm0.0198$ & $0.739$ 
&& $0.732\pm0.027$ & $0.730$
\\
\cline{1-1} \cline{3-4} \cline{6-7}
$-2\ln {\cal L}$ & & \multicolumn{2}{|c|}{$8140.06$}
& & \multicolumn{2}{|c|}{$7608.39$}
 \\
\cline{1-1} \cline{3-4} \cline{6-7}
\end{tabular}
\caption{\footnotesize Power law (PL) parameter results:  means, standard deviations (left subdivision of columns) and maximum likelihood values (right subdivision of columns)  with CMB data  (WMAP7 $+$ BICEP $+$ QUAD) and all data ($+$UNION2 $+ H_0 +$ BBN) in a flat universe. $H_0$ and $\Omega_\Lambda$ constraints are derived from the other parameters.}
\label{table:alldata_PL_flat}
\end{table*}

\begin{table*}[tbp]
\centering
\begin{tabular}{|l|r@{}|c|c|r@{}|c|c|r@{}|c|c|}
\hline
Parameters & & \multicolumn{2}{|c|}{All Data $I_{1,\rm max}=1/\sqrt{2}$} & & \multicolumn{2}{|c|}{All Data $I_{1,\rm max}=1/2$}
& & \multicolumn{2}{|c|}{CMB Only $I_{1,\rm max}=1/\sqrt{2}$}
\\
\cline{1-1} \cline{3-4} \cline{6-7} \cline{9-10}
$100\Omega_b h^2$  & & $2.279\pm0.107$ & $2.227$ & & $2.282\pm0.107$ & $2.410$
& & $2.160\pm0.159$ & $2.110$ \\
$\Omega_c h^2$     & & $0.1127\pm0.0055$ & $0.1101$ & & $0.1126\pm0.0056$ & $0.1100$ 
& & $0.1297\pm0.0142$  & $0.1338$\\
$\theta$           & & $1.0411\pm0.0030$ & $1.0402$ & &  $1.0411\pm0.0030$  & $1.0417$
 & & $1.0395\pm0.0032$ & $1.0381$  \\
$\tau$             & & $0.086\pm0.016$ & $0.096$ & & $0.088\pm0.016$ & $0.091$ 
 & & $0.082\pm0.016$ & $0.072$ \\
$\bar{G^{\prime}}$  & & $0.0122\pm0.0268$ & $0.0055$ && $0.0191\pm0.0248$ & $0.0213$
& & $0.0186\pm0.0283$ & $0.0221$ \\
$\ln[10^{10}A_c]$   & & $0.0032\pm0.0117$ & $0.0036$ && $0.0032\pm0.0122$ & $0.0098$
& & $0.0051\pm0.0122$  & $0.0056$ \\
$m_1$&& $0.0048\pm0.0073$ & $0.0060$ & &  $0.0025\pm0.0071$ & $0.0068$
&&  $0.0021\pm0.0078$  & $0.0009$ \\
$m_2$&& $0.0152\pm0.0122$ & $0.0163$ & &  $0.0120\pm0.0122$ & $0.0109$
&&  $0.0086\pm0.0137$  & $0.0104$ \\
$m_3$&& $-0.0120\pm0.0181$ & $-0.0042$ & & $-0.0140\pm0.0179$ & $-0.0085$
&& $-0.0151\pm0.0191$  & $-0.0161$\\
$m_4$&& $0.0427\pm0.0190$ & $0.0460$ & & $0.0327\pm0.0171$ &  $0.0481$ 
&& $0.0455\pm0.0195$  & $0.0583$\\
$m_5$&& $0.0198\pm0.0256$ & $0.0050$ & & $0.0168\pm0.0249$ & $0.0486$
&& $0.0165\pm0.0272$  & $0.0165$\\
$m_6$&& $-0.0156\pm0.0325$ & $-0.0089$ & & $-0.0142\pm0.0328$ & $-0.0166$
&& $0.0062\pm0.0377$ & $-0.0120$\\
$m_7$&& $-0.0061\pm0.0354$ & $-0.0015$ & & $-0.0060\pm0.0333$ & $-0.0060$
&& $-0.0174\pm0.0383$  & $-0.0324$\\
$m_8$&& $0.0278\pm0.0486$ & $0.0285$ & & $0.0403\pm0.0464$ & $0.0431$
&&  $0.0174\pm0.0505$  & $-0.0061$\\
$m_9$&& $-0.1239\pm0.0731$ &  $-0.1436$ & & $-0.0970\pm0.0670$ & $-0.1458$
&& $-0.1319\pm0.0770$  & $-0.1184$\\
$m_{10}$&& $0.0336\pm0.0609$ & $0.0219$ & & $0.0282\pm0.0602$ & $0.0462$
&& $0.0150\pm0.0647$  & $0.0441$ \\
$m_{11}$&& $0.0759\pm0.0908$ & $0.0225$ & & $0.0599\pm0.0847$ & $0.0364$
&& $0.0591\pm0.0966$  & $0.1339$\\
$m_{12}$&& $-0.0917\pm0.1027$ & $-0.1604$ & & $-0.0702\pm0.0946$ & $-0.1477$
&& $-0.1100\pm0.1076$  & $-0.2137$\\
$m_{13}$&& $-0.0947\pm0.1129$ & $-0.1895$ & & $-0.0764\pm0.1036$ & $-0.1577$
&& $-0.0506\pm0.1194$  & $-0.2300$ \\
$m_{14}$&& $0.1116\pm0.1616$ & $0.2069$ & & $0.0561\pm0.1450$ &  $0.2126$ 
&& $0.1507\pm0.1714$ & $0.2103$\\
$m_{15}$&& $-0.0199\pm0.2042$ & $0.0617$ & & $0.0191\pm0.1864$ & $-0.0091$
&& $-0.0255\pm0.2152$ & $0.0686$ \\
$m_{16}$&& $0.1006\pm0.0975$ & $0.1318$ & & $0.0837\pm0.0964$ & $0.1102$ 
&& $0.1481\pm0.1043$  & $0.0772$\\
$m_{17}$&& $-0.1253\pm0.2688$ & $-0.1953$ & & $-0.1094\pm0.2326$ & $-0.1302$ 
&& $-0.0575\pm0.2833$  & $-0.1376$ \\
$m_{18}$&&  $-0.5089\pm0.2938$ & $-0.6131$ & & $-0.3322\pm0.2475$ & $-0.3798$ 
&& $-0.4894\pm0.3083$  & $-0.6610$\\
$m_{19}$&& $0.2239\pm0.3773$ & $0.2737$ & & $0.1524\pm0.3028$ & $0.1785$ 
&& $0.2406\pm0.3878$ & $0.5228$ \\
$m_{20}$&& $-0.0742\pm0.4070$ & $0.0011$ & & $-0.2472\pm0.3173$ & $-0.1789$ 
&& $-0.1265\pm0.4065$ & $-0.0113$\\
\cline{3-4}\cline{6-7} \cline{9-10}
$n_s$              & & $1.0299\pm0.0671$ &  $1.1296$ & & $1.0075\pm0.0515$ & $1.0535$ 
& & $1.0191\pm0.0672$  & $1.0823$\\
$\ln[10^{10}A_s]$  & & $3.0387\pm0.0582$ & $3.0358$ & & $3.0446\pm0.0573$ & $3.0654$ 
& & $3.0684\pm0.0626$  & $3.0726$\\
$H_0$              & & $71.03\pm2.28$ & $71.22$  & &  $71.08\pm2.28$ & $73.28$ 
& & $63.86\pm5.88$  & $61.35$ \\
$\Omega_\Lambda$ && $0.730\pm0.026$ & $0.739$ & & $0.731\pm0.026$ & $0.750$
&&   $0.614\pm0.105$ & $0.588$  \\
\cline{1-1} \cline{3-4}\cline{6-7}  \cline{9-10}
$2\Delta \ln L$ & & \multicolumn{2}{|c|}{$16.85$} && \multicolumn{2}{|c|}{$14.26$}  & & \multicolumn{2}{|c|}{$17.2$} \\
\cline{1-1} \cline{3-4} \cline{6-7} \cline{9-10}
\end{tabular}
\caption{\footnotesize 20 principal component (PC) parameter results: means, standard deviations (left subdivision of columns) and maximum likelihood (ML) values (right subdivision of columns).   Fiducial results are for all data and nonlinearity prior $\Imax=1/\sqrt{2}$, left columns, with variations shown in
center and right columns.   Parameters $n_s-\Omega_\Lambda$ are derived from the chain parameters. The difference in likelihood $2\Delta \ln L$ is given for the ML values and taken with respect to the corresponding PL maximum likelihood model in Tab.~\ref{table:alldata_PL_flat}.}
\label{table:alldata_20PCs_flat_priorI1}
\end{table*}

For the WMAP7 data \cite{Larson:2010gs}, we use the optimized approximate likelihood 
from Ref.~\cite{DvoHu10a}. 
In addition, we utilize data from BICEP and QUAD which include polarization constraints \cite{Chiang:2009xsa,Brown:2009uy}.
We calculate the CMB power spectra with gravitational lensing turned off and 
the default sparse sampling in $\ell$ (accuracy\_boost=1).  We correct for these approximations in postprocessing
by importance sampling as described in Appendix \ref{sec:MCMCopt} before presenting the
results in the next section.  The main effect is a $\sim 0.5\sigma$ upwards shift in the
$\Omega_b h^2$ posterior to compensate the smoothing effect of lensing.

In order to ensure that models are compatible with a reasonable cosmology we 
add non-CMB constraints from
the 
UNION2 supernovae dataset \footnote{\url{http://www.supernova.lbl.gov/Union}}, the SHOES $H_0= (74.2 \pm 3.6)$ km/s/Mpc measurement \cite{SHOES}
and a big bang nucleosynthesis constraint of $\Omega_b h^2 = 0.022 \pm 0.002$.
These data mainly constrain the energy density components of the universe rather
than the inflationary initial conditions.  We call the combination of CMB and external
data the ``all data" analysis.  We address the impact of the $\Imax$ prior and the non-CMB
data in \S \ref{sec:robustness} below.

\begin{figure}[tbp]
\includegraphics[width=0.475\textwidth]{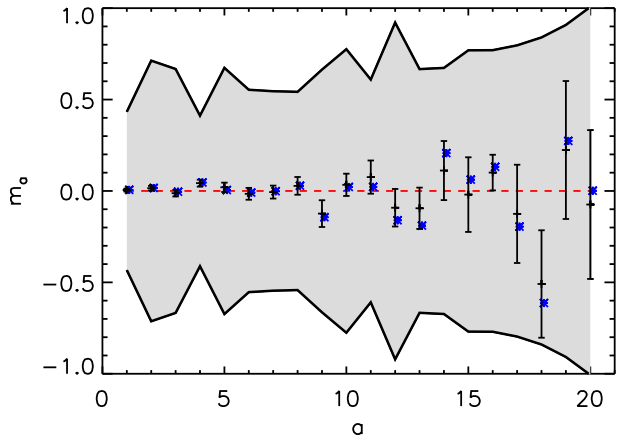}
\caption{Constraints on the 20 PC amplitudes   from the all-data analysis with an
$\Imax < 1/\sqrt{2}$ prior.  The only significant deviation from the $m_a=0$ PL
expectation (red dashed line) is  $m_4 = 0.0427\pm0.0190$.  The impact of the prior can be visualized by taking the maximum amplitude of an individual $m_a$ that satisfies
the prior (solid lines), which implies that only  $m_{17}-m_{20}$  are significantly prior limited.   The maximum likelihood
model is shown as starred points.}
\label{plot:ma_data}
\end{figure}

\section{MCMC Results}
\label{sec:results}

In this section, we present the results of the Markov Chain Monte Carlo (MCMC)
analysis in the
principal component (PC) space of the GSR source function.   We discuss the results of
our fiducial all data analysis in \S \ref{sec:alldata} and address the impact of priors
and non-CMB data in \S \ref{sec:robustness}.

\subsection{All Data}
\label{sec:alldata}

For our fiducial results we use the all-data combination of CMB and external
data described in the previous section.  To establish a baseline for the 
PC results we start with the $m_a=0$ power law (PL) case, $\Delta_{\cal R}^2=A_s(k/k_p)^{n_s-1}$.
Table \ref{table:alldata_PL_flat} gives the mean, standard deviation of the posterior probabilities, and the maximum likelihood (ML) parameter values for the power law model.

For the PC analysis, we take 20 components and a nonlinearity tophat prior of $\Imax<1/\sqrt{2}$ (see \S \ref{sec:GSRpc}).  
Table \ref{table:alldata_20PCs_flat_priorI1} gives the parameter constraints
as well as the maximum likelihood PC model (left columns).

\begin{figure}[btp]
\includegraphics[width=4in]{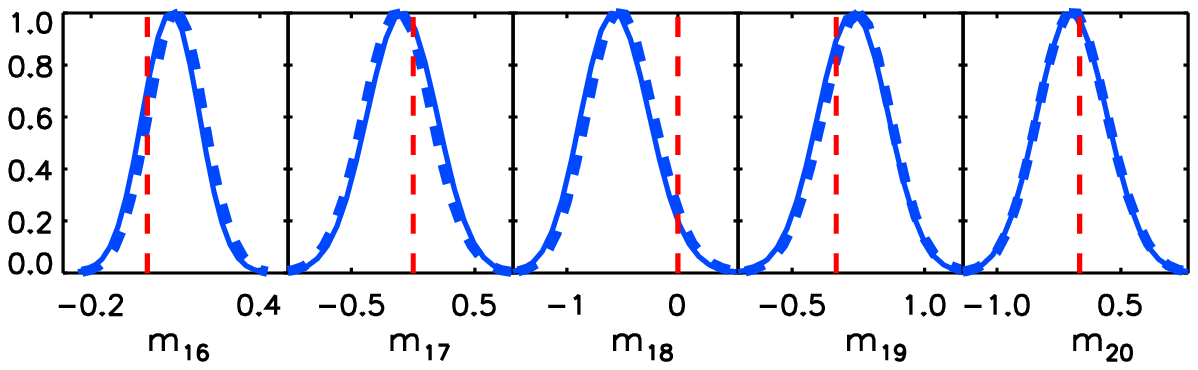}
\includegraphics[width=4in]{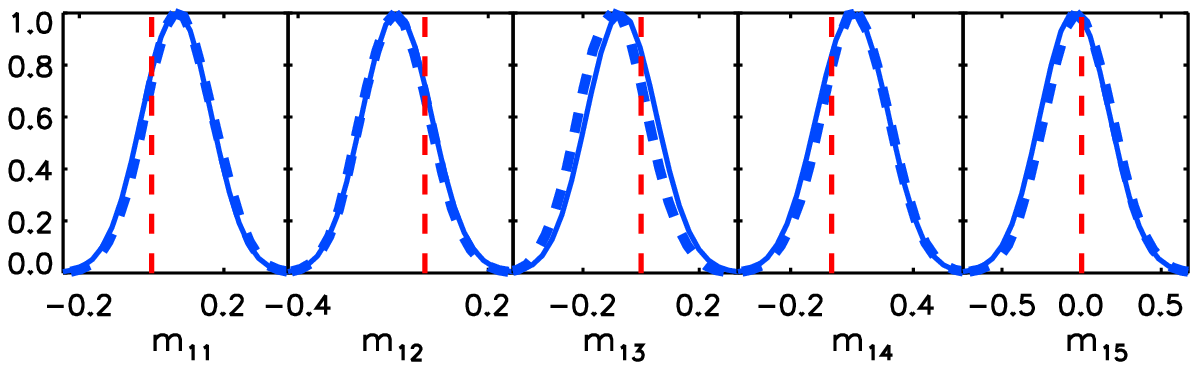}
\includegraphics[width=4in]{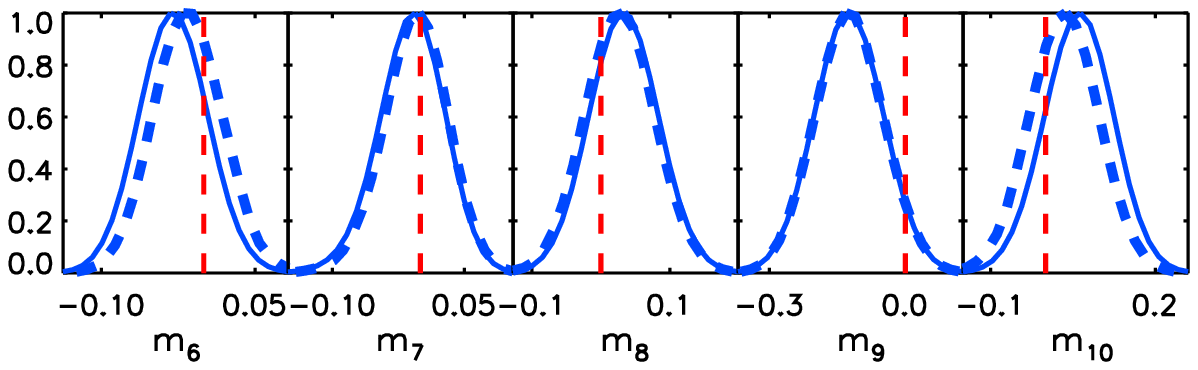}
\includegraphics[width=4in]{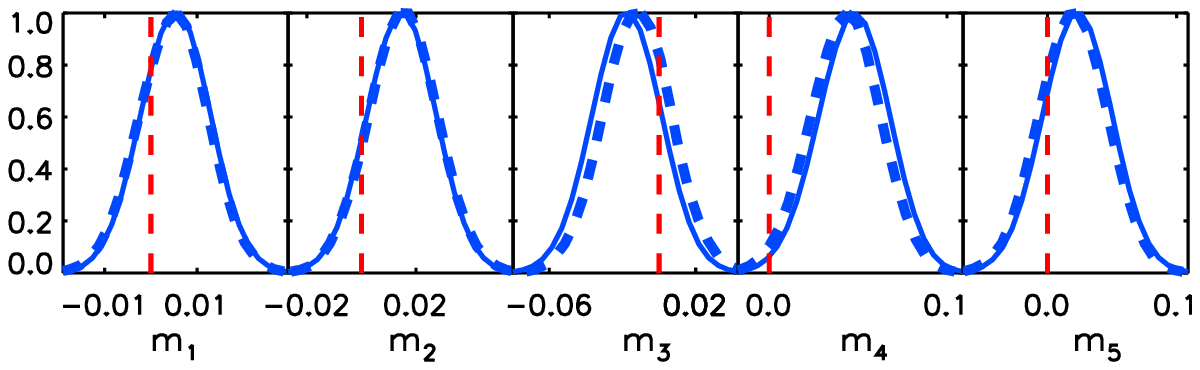}
\includegraphics[width=4in]{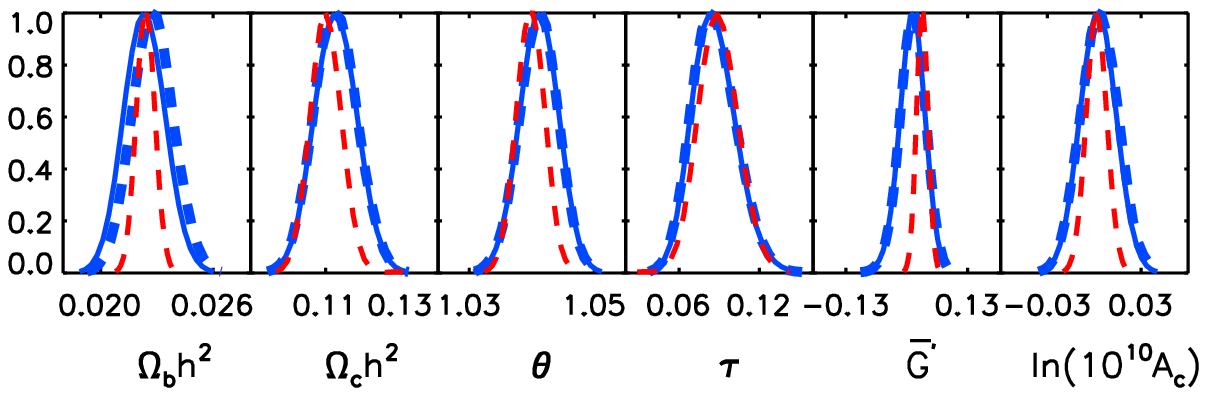}
\caption{\footnotesize Parameter probability distributions from the all-data analysis in 
a flat universe with $\Imax=1/\sqrt{2}$.    Dashed lines represent the posteriors with
approximations for the low $\ell$ polarization likelihood and $C_\ell$ accuracy used
to run the MCMC (see Appendix); solid lines represent posteriors corrected by importance
sampling.  Red dashed lines represent corrected posteriors for power law models. }
\label{plot:alldataposteriors}
\end{figure}

The improvement in the ML PC  model over the ML power law model is $2\Delta \ln {\cal L}= 17$ for
20 extra parameters and so is not statistically significant in and of itself.
Of course, specific inflationary models may realize this improvement with a smaller set of
physical rather than phenomenological parameters (see \S \ref{sec:model}),  and so it is interesting to examine more closely the origin of this
improvement.  

The main improvement comes from $\ell \le 60$ in the TT part of the WMAP likelihood with a  $2\Delta \ln {\cal L}=11.9$.
We shall see in \S \ref{sec:modelindependent} that these improvements are largely associated with known features in the WMAP temperature power spectrum.

In terms of the principal components, the improvements are localized in only a few
of the 20 parameters. 
Fig.~\ref{plot:ma_data} plots these $m_a$ constraints and ML values.  
Most of the components are consistent with zero at the $\sim 1\sigma$ level.
Components $m_{17}-m_{20}$ are constrained in part by the $\Imax$ prior not just
the data.

\begin{figure*}[tbp]
\includegraphics[width=0.45\textwidth]{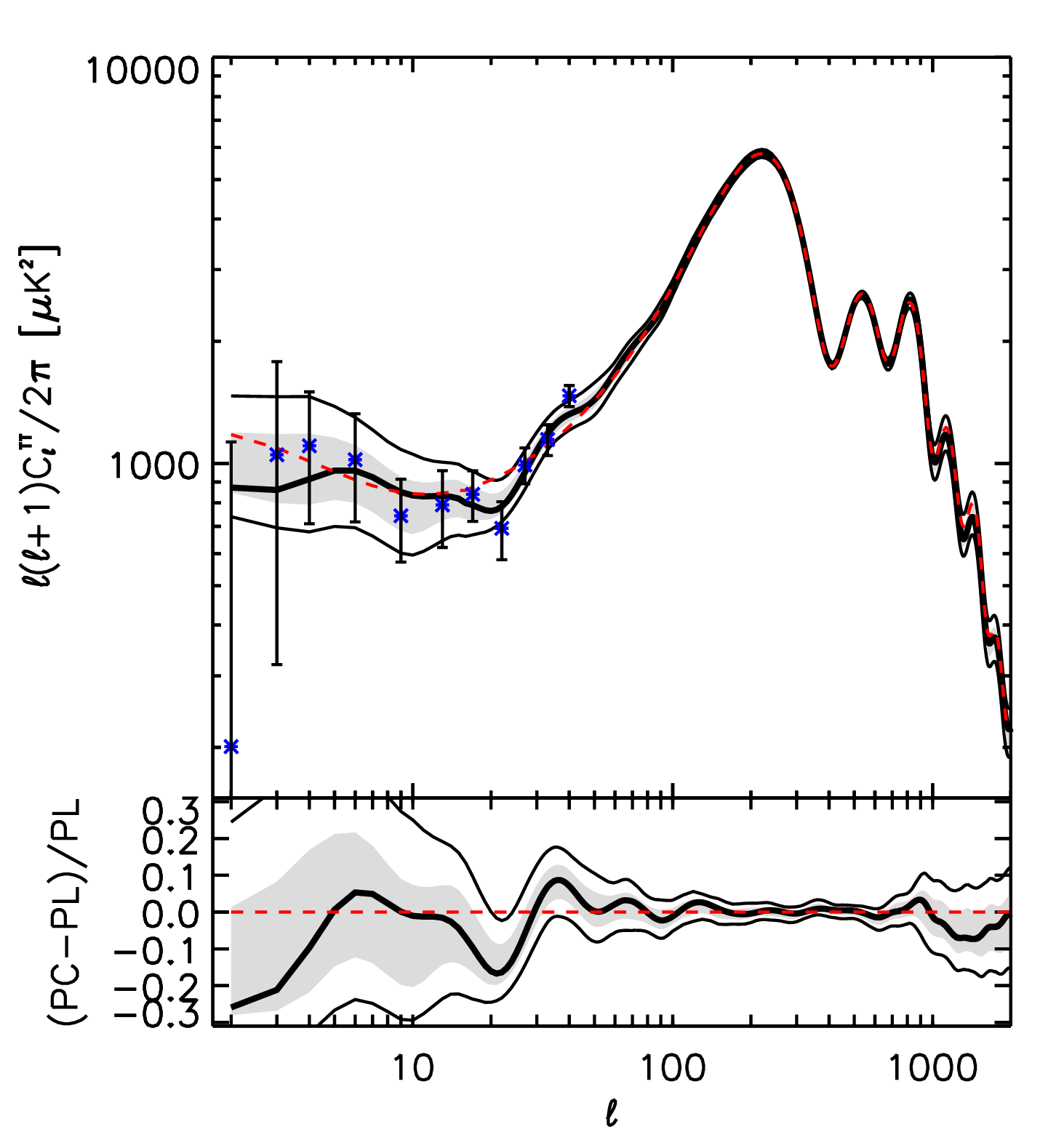}
\includegraphics[width=0.45\textwidth]{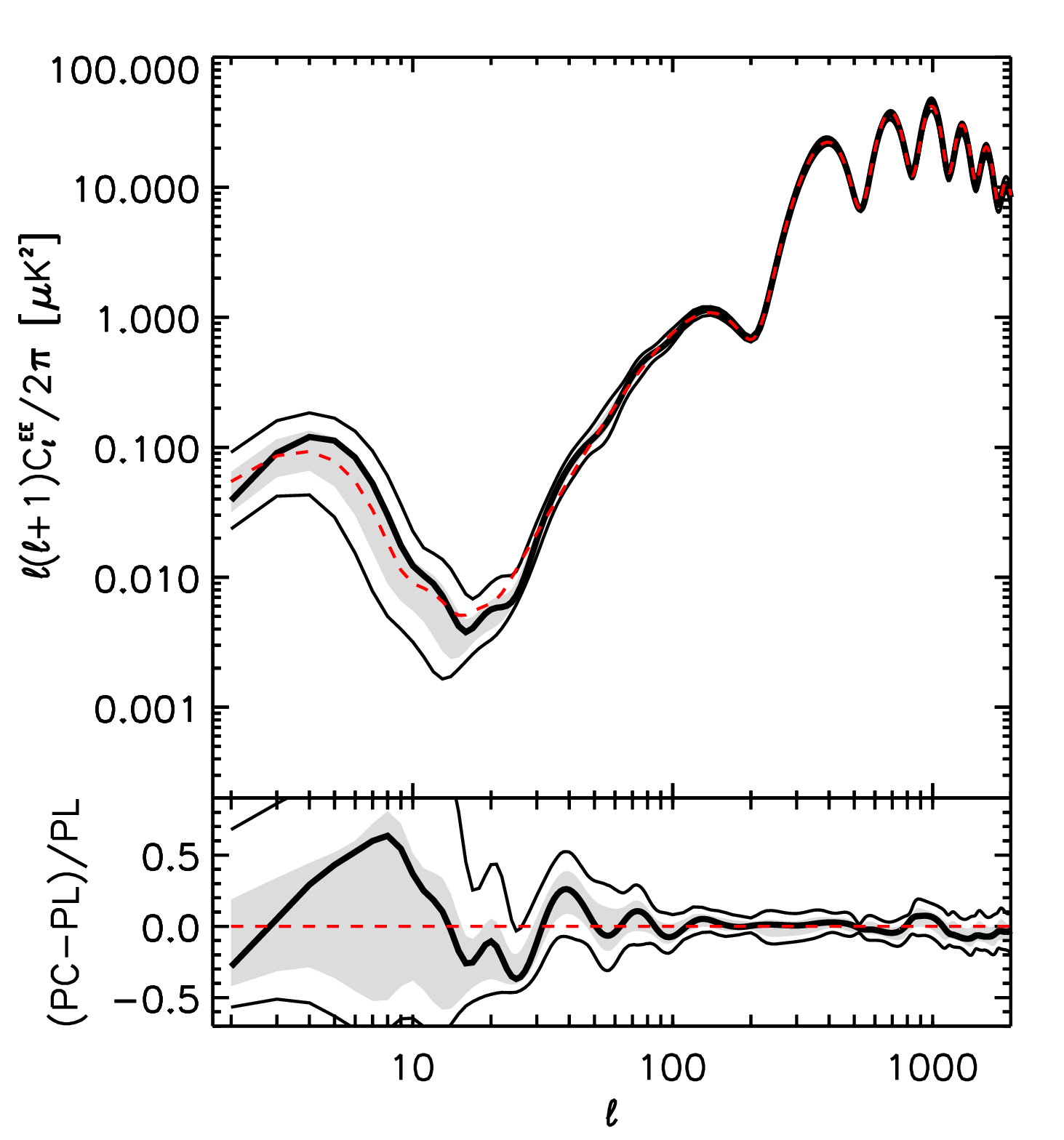}
\caption{The temperature (left) and $E$-mode polarization (right) power spectra posterior using the all-data PC constraints and a prior of $I_{1,\rm max}=1/\sqrt{2}$. The shaded area encloses the $68\%$ CL region and the upper and lower curves show the upper and lower $95\%$ CL limits. The maximum likelihood (ML) model is shown as the thick black central curve, and the power law ML model is shown in red dashed lines. The blue points with error bars show the $7$-year WMAP measurements.}
\label{plot:Clposterior}
\end{figure*}

As in the 5 PC analysis of \cite{DvoHu10a}, the single most discrepant parameter
between the PL and PC cases is $m_4$ corresponding to a feature centered around
$\eta \sim 300$Mpc and resembling a local running of the tilt.
Fig.~\ref{plot:alldataposteriors} shows the posterior probability distributions of the
parameters.  An $m_4$ value as extreme as the power
law value of $m_4=0$ is disfavored at $98.2\%$ CL compared with $94.8\%$ for 5 PCs and
WMAP7 alone.   The increase in significance by a fraction of a $\sigma$
arises because of the correlation between $m_4$ and the higher principal components. 
Perhaps more importantly, freedom in the higher PCs allows large $m_4$ without the
need to make large adjustments to the cosmological parameters that would violate non-CMB constraints.  On the other hand, one event out of 20 showing a $98\%$ exclusion is not
that unlikely.

 \begin{figure}[tbp]
\includegraphics[width=0.475\textwidth]{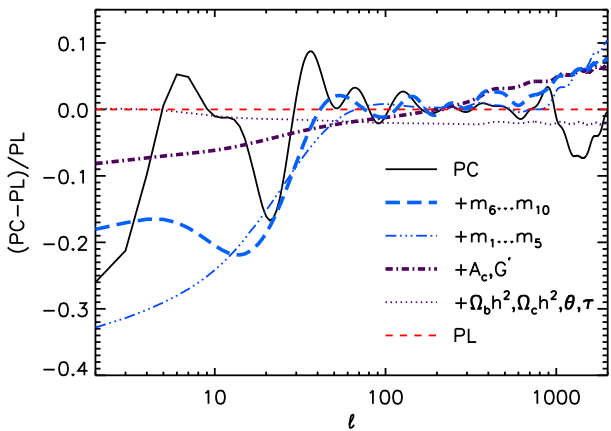}
\caption{Parameter decomposition of the temperature power spectrum 
difference between the power law (PL) and PC maximum likelihood (ML) models shown in Fig.~\ref{plot:Clposterior} (bottom panel).   The curves
include cumulative changes in parameters between the models starting
with the cosmological parameters,
adding the normalization $A_c$ and effective tilt $\bar G'$,  $m_1 \ldots m_5$, etc. until the full PC ML parameters are
utilized.}
\label{plot:ClTT_20PCs_ML_alldata_I1prior1oversqrt2_parameter_decomposition}
\end{figure}

The poorly constrained $a>10$ modes allow large amplitude deviations and in fact even marginally prefer them.   This explains why including the higher components can change
results on the lower components.   
Large amplitude deviations in the high order components make the modes no longer statistically independent as they would be for infinitesimal deviations.    Still the correlation remains relatively small.   For example $R_{4a}= {\rm Cov}(m_4,m_a)/\sigma_{m_4}\sigma_{m_a}$ reaches $0.4$ only for one mode, $m_5$, with more typical correlations in the $\pm 0.1-0.2$ range.

The next most significant deviations are in $m_9$ (with a value of $m_9 = 0$ disfavored at the $89.6\%$ CL) and $m_{18}$ (with a value of $m_{18} = 0$ disfavored at the $91.8\%$ CL). These results are also consistent with the PL null hypothesis of $m_a=0$, given that there are only $3$ events out of $20$ where tests of that model exceed the $\sim 90\%$ CL.

We can get further insight on the origin of these constraints by examining
the  maximum likelihood (ML) models.
Fig.~\ref{plot:Clposterior} show temperature and polarization power spectra of
the ML PL (red dashed lines) and PC (thick solid curve) models respectively.  The poorly constrained
$a>10$ modes create fluctuations in the low order multipoles which marginally fit
features in the data better such as the low quadrupole and glitch at $\ell \sim 20-40$. 
These large amplitude modes require small amplitude low order PC variations in order
to compensate the broad band residual effects they have.
 This can be seen by decomposing the
difference between the ML PL and PC models into contributions from the various parameters (see Fig.~\ref{plot:ClTT_20PCs_ML_alldata_I1prior1oversqrt2_parameter_decomposition}).
Removing the large $m_{10}-m_{20}$ components from the model not only removes the low $\ell$ oscillations but also creates broadband deviations, especially at $\ell \lesssim 40$,  that are compensated by a combination of small amplitude changes in $m_{1}-m_{5}$ and 
effective tilt $\bar G'$.

\subsection{Robustness Tests}
\label{sec:robustness}

In order to test the robustness of the fiducial results of the last section, we run separate MCMC chains with different choices for
the nonlinearity prior and data sets.

\begin{figure}[tbp]\includegraphics[width=0.475\textwidth]{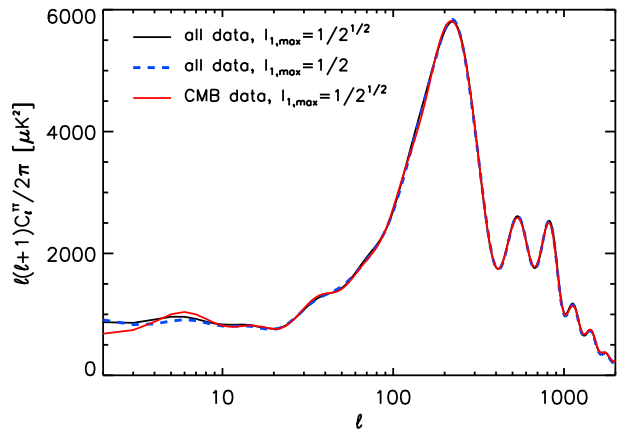}
\caption{Comparison of the maximum likelihood models of the three
MCMCs of Tab.~\ref{table:alldata_20PCs_flat_priorI1}:
 the all-data analysis with $\Imax=1/\sqrt{2}$ (black curve), all-data with $\Imax=1/2$ (blue curve), and CMB data with $\Imax=1/\sqrt{2}$ (red curve).  The smallness of the differences indicates robustness of our results to the priors and external data sets.}
\label{plot:ClTT_20PCs_ML_models}
\end{figure}

\begin{figure}[tbp]
\includegraphics[width=0.475\textwidth]{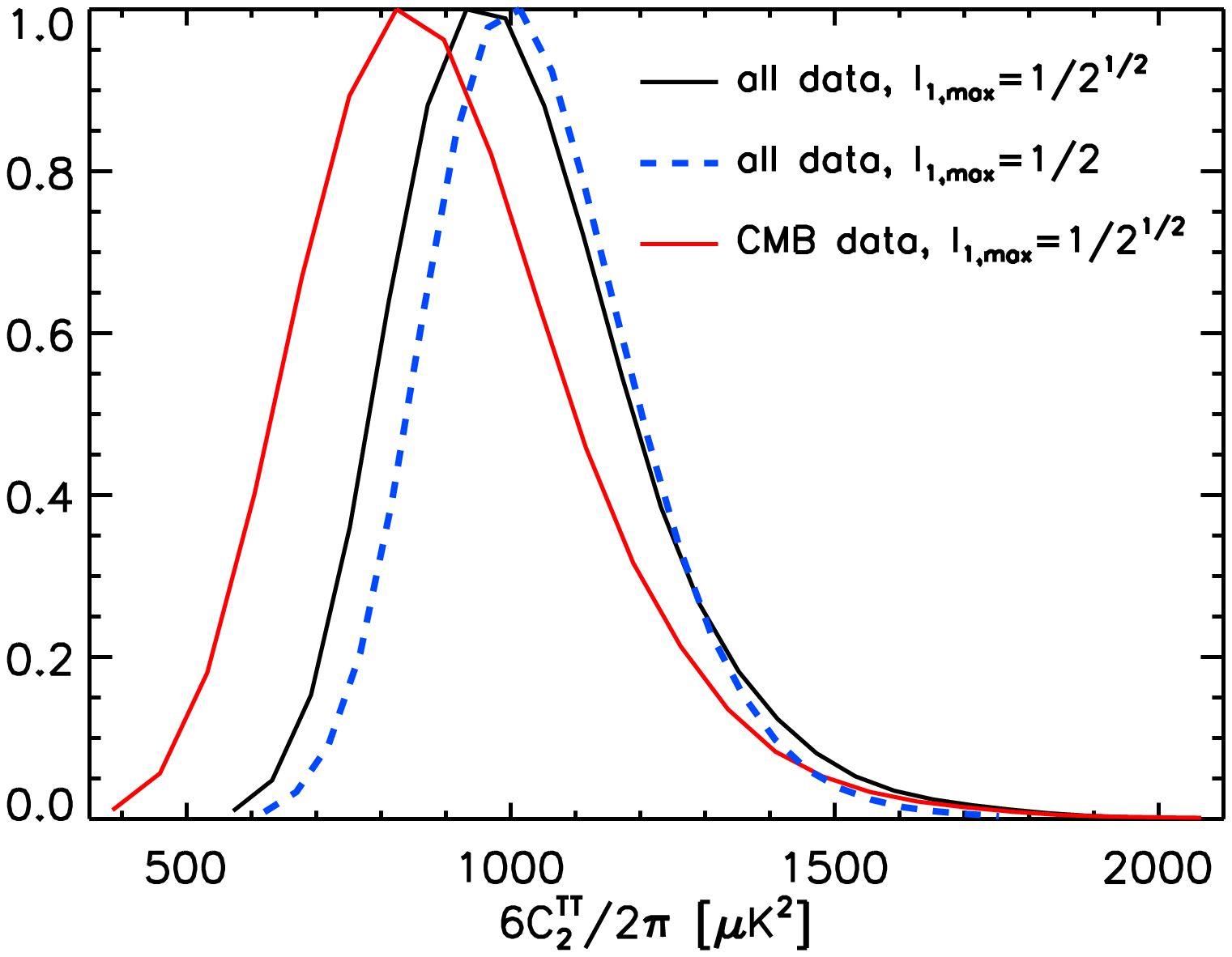}
\caption{The temperature quadrupole power $C_2^{TT}$  distribution for the all-data analysis with $\Imax=1/\sqrt{2}$ (black curve), all-data with $\Imax=1/2$ (blue curve), and CMB data with $\Imax=1/\sqrt{2}$ (red curve).  Without external data to constrain the
cosmological constant, the quadrupole can be lowered by reducing the integrated Sachs-Wolfe effect.}
\label{plot:C2TT_distrib}
\end{figure}

We first examine the impact of our $\Imax$ prior by reanalyzing the all-data case with $\Imax=1/2$ instead of $1/\sqrt{2}$ (see Tab.~\ref{table:alldata_20PCs_flat_priorI1}).
The main impact of tightening the prior is on $m_{18}-m_{20}$ as is expected from
Fig.~\ref{plot:ma_data}.
 These components
mainly affect the low $\ell$ multipoles.   In spite of this fact the prior on $\Imax$ has very little impact on the behavior of favored models at low $\ell$.   
In Fig.~\ref{plot:ClTT_20PCs_ML_models}, we show the maximum likelihood model
with the stronger $\Imax$ prior.    Even at low $\ell$ the differences are much smaller
than cosmic variance.   In particular 
the posterior distribution of power in the quadrupole moment for models in the chain shown in Fig.~\ref{plot:C2TT_distrib} differ negligibly.  

Some of this robustness in the low multipole moments is due to the impact of the
non-CMB data.   
 Without the external data,
 the quadrupole distribution extends to smaller quadrupole moments
due to the ability to reduce the integrated Sachs-Wolfe effect by lowering the
cosmological constant in the absence of constraints on the acceleration of the expansion
(see Fig.~\ref{plot:C2TT_distrib}).   In this case the data may prefer more extreme inflationary models that further lower the quadrupole that are excluded by our nonlinearity prior
on $\Imax$ \cite{Contaldi:2003zv}.

 The main impact on parameters of removing
the non-CMB data is to allow a wider range in $\Omega_c h^2$ (see Table \ref{table:alldata_20PCs_flat_priorI1}).   In contrast to the
5PC analysis \cite{DvoHu10a}, this wider range though has little impact on the PC parameters.   In particular the higher order PC components allow compensation of
the effects of $m_4$ across the acoustic peaks without the need to vary $\Omega_c h^2$
substantially.    For similar reasons, we expect our flatness prior to have little impact on
the PC results aside from weakening the constraints on $\Omega_\Lambda$
and $\Omega_c h^2$ and small shifts of the location of features in $G'$ with the angular diameter distance
degeneracy.

\section{Applications}
\label{sec:apps}

Here we discuss applications of the fiducial 20 PC analysis of \S \ref{sec:alldata}.
In \S \ref{sec:modelindependent} we  place
constraints on and devise tests of slow roll and single field inflation in a model independent manner.   Alternately, as a complete observational basis for efold bandlimited models, 
the PC analysis places constraints on any such model that 
 satisfies
the GSR condition.  We use  running of the tilt and a step in the inflaton potential as
example test cases 
 in \S \ref{sec:model}.

\subsection{Testing Slow Roll and Single Field Inflation}
\label{sec:modelindependent}

Bounds on the PC components can be thought of as functional constraints on 
$G'$ itself across the observed range from WMAP.  These in turn limit features in the inflaton
potential $V(\phi)$ through the approximate relation \cite{Dvorkin:2009ne}
\ba
G'(\ln\eta) &\approx& 3 ({V_{,\phi}\over V})^2 - 2{V_{,\phi\phi}\over V}  \,.
\ea
If the inflaton carries non-canonical kinetic terms then the relationship is modified
to include variations in the sound speed \cite{Hu11}.

Since the PC decomposition only represents features in $G'$ across the observable domain, one
should consider the constraints on the $m_a$s as defining a PC filtered version of $G'$:
\begin{equation}
G'_{20}(\ln \eta) = \sum_{a=1}^{20} m_a S_a(\ln\eta) \,.
\label{eq:G20}
\end{equation} 
Any significant deviation from zero of this function would indicate a violation of ordinary slow
roll.  
We can extract the posterior probability of $G'_{20}$ by considering its values
on a continuous set of samples of $\eta$ as
derived parameters.

\begin{figure}[tbp]
\includegraphics[width=0.45\textwidth]{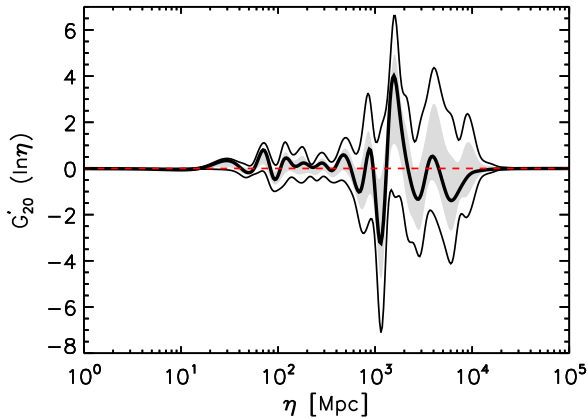}
\caption{The $20$ PC filtered $G'$ posterior from  the fiducial all-data analysis
and $I_{1,\rm max}=1/\sqrt{2}$ as a prior. The shaded area encloses the $68\%$ CL region and the upper and lower curves show the upper and lower $95\%$ CL limits. The maximum likelihood is shown as the thick black central curve, and the power law ML model is shown in red dashed lines.}
\label{plot:Gprime_posterior_WMAP7_BICEP_QUAD_20PCs_SN_H0_BBN_flat}
\end{figure}

In Fig. \ref{plot:Gprime_posterior_WMAP7_BICEP_QUAD_20PCs_SN_H0_BBN_flat} we 
plot both the ML model and the 68\% and 95\% posterior bands.     Note that $G'_{20}=0$ lies within the 95\% CL regime for all $\eta$.
These functional constraints differ from a full reconstruction of $G'$ in that the PCs
filter out deviations at $\eta < 20$ Mpc and $\eta > 10^4$ Mpc as well as deviations that are
too high frequency to satisfy our bandlimit.

 In the well-constrained regime of $30 \lesssim \eta/{\rm Mpc} \lesssim 400$ constraints
are both tight and  consistent with $G'_{20}=0$.  
  Only nearly zero mean high frequency deviations
are allowed in this regime.  
 Nonetheless, the poorly constrained $m_{10}-m_{20}$ components allow, but do not strongly prefer, large
oscillatory features between
$10^3 \lesssim \eta/{\rm Mpc} \lesssim 10^4$.  
In fact $G'_{20}=0$ lies noticeably outside the
68\% CL bands only for the dip and bump between $1000-2000$ Mpc and a bump at $70-100$ Mpc. 

We can associate the most significant features with the corresponding effects on the 
observable power spectra themselves.  
Figure~\ref{plot:Clposterior}
shows the 68\% and 95\% range in the power spectra posterior. 
The $1000-2000$ Mpc feature in fact corresponds to the $\ell=20-40$ dip and bump in the temperature power spectrum.  
The $70-100$ Mpc feature corresponds to a glitch at $\ell \sim 600-700$ \cite{Covi:2006ci,Hamann:2007pa}.  
While the $\eta \gtrsim 10^4$ Mpc regime is limited by our priors on the amplitude of deviations through $\Imax$ we have shown that the data do not favor a feature corresponding to a low quadrupole $\ell=2$ unless acceleration constraints are omitted
(see \S \ref{sec:robustness}).

Finally, we can examine the posterior distributions of the $E$-mode polarization.  These predictions are not
significantly constrained by the polarization data sets employed. 
Instead these distributions are limited mainly by the common origin of the temperature and polarization spectra from
single field inflation.
   These serve as predictions for
future measurements.  
For example, the low significance features in the temperature power spectrum
predict corresponding ones in the $E$-mode polarization which have yet to be measured
and can be used to test the hypothesis of their inflationary origin at substantially higher
joint significance \cite{Mortonson:2009qv}. 
 In particular, one expects  a 
$\sim 26\% ^{+13\%}_{-17\%}$  enhancement in the $EE$ power spectrum at $\ell = 39$ and a $\sim -37\% ^{+17\%}_{-3\%}$ deficit around $\ell =25$.   The skew distribution in the latter case reflects the difficulty in constructing models with low power out of the principal components rather than the data disfavoring such models.   Models that actually explain the low $TT$ power at $\ell=25$  predict low $EE$ power as well.

Even in the acoustic regime where the polarization predictions are tight and do not suggest the presence of features,   these predictions are of interest.  If
future observations violate them, then not only will slow-roll inflation be falsified but
all single field inflationary models,   including those with sound speed variations, as long as they satisfy our weak prior constraint on acceptable models: the 
efold bandlimit and small GSR non-linearity $\Imax < 1/\sqrt{2}$.
Such a violation might indicate other degrees of freedom breaking the relationship between
the temperature and polarization fields, e.g. isocurvature modes in multifield inflation or trace amounts of cosmological defects.
For $\ell \lesssim 30$ violation could alternately indicate a more complicated reionization
scenario  \cite{Mortonson:2009qv}.

Currently these bounds and tests apply to the  $\ell < 800$ regime measured by WMAP but will soon be extended
by high resolution ground based experiments and Planck.

\subsection{Constraining Inflationary Models}
\label{sec:model}

We can also apply the model independent principal component analysis to any specific set of models
that satisfy the GSR condition $\Imax < 1/\sqrt{2}$ and  bandlimit of features no sharper than about 1/4 efold.  To place constraints
on the parameters of a model, one projects the source function $G'$ of the model onto the principal 
components
\begin{equation}\label{eq:mas}
m_a = {1 \over \ln\eta_{\rm max}-\ln\eta_{\rm min}} \int_{\eta_{\rm min}}^{\eta_{\rm max}} {d\eta\over \eta} S_a(\ln\eta) G'(\ln\eta)
\end{equation}
 as a function of parameters and compares the result to the joint posterior probability distributions of the components.
 Likewise one can construct $G'_{20}$ from the result and compare it with Fig.~\ref{plot:Gprime_posterior_WMAP7_BICEP_QUAD_20PCs_SN_H0_BBN_flat}.

 In fact, the means and
covariance matrix ${\bf C}$ of the components $m_a$ form a simple but useful representation of the joint PC posteriors.   From these, one can construct a $\chi^2$ statistic
\begin{equation}
\chi^2 = \sum_{a,b=1}^{20} \left[ (m_a-\bar m_a) {\bf C}_{ab}^{-1}  (m_b-\bar m_b) \right]\,,
\end{equation}
or the likelihood ${\cal L} \propto \exp(-\chi^2/2)$  under a multivariate Gaussian approximation to the posteriors.   For example the ML PC model gives an improvement of
$\Delta \chi^2 = -15.36$ over PL to be compared with $-2\Delta \ln L = -16.85$.

As a simple illustration of a concrete model, consider a linear deviation in $G'$
\begin{equation}\label{eq:model_running}
G'(\ln\eta) = 1- n_0  + \alpha \ln\left( \eta/\eta_0\right) \,.
\end{equation}
The curvature power spectrum for this model has a local tilt of 
\begin{eqnarray}
{d \ln \Delta^2_{\cal R} \over d \ln k}= n_0-1 + \alpha \ln\left({k\eta_0 \over C}\right) - {\alpha \pi \over \sqrt{2}}{I_1 \over 1+ I_1^2}\, ,
\end{eqnarray}
where $C = e^{7/3-\gamma_E}/2 \approx 2.895$ and 
\begin{equation}
I_1 = {1 \over \sqrt{2} } \left[ {\pi \over 2} (1-n_0 -\alpha\ln k\eta_0)+ 1.67\alpha \right].
\label{eqn:alpharunning}
\end{equation}
For $|n_0-1| \ll 1$ and $|\alpha| \ll 1$, the $I_1$ term contributes negligibly and the model gives
a linear running of the tilt \cite{DvoHu10a}.  

\begin{figure}[tbp]
\includegraphics[width=0.45\textwidth]{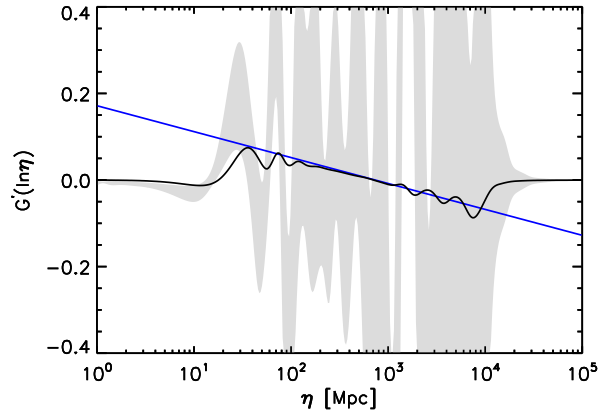}
\caption{A model with a linear deviation in $G'$ with slope $\alpha=-0.026$ (and arbitrary offset) is shown as the blue curve. The $20$ PC filtered source $G_{20}^{\prime}$ (in black lines) is compared with the input linear $G'$ model. $20$ PCs captures all of the observable information in $\alpha$.  These models are compatible with the 68\% CL region (shaded) for $G_{20}'$ from the fiducial all data analysis.}
\label{plot:Gprime_filtered_WMAP7_20_vs_50PCs}
\end{figure}

The 20 PC components are a linear 
function of $\alpha$ given explicitly by
\begin{equation}
m_a(\alpha) ={\alpha \over \ln\eta_{\rm max}-\ln\eta_{\rm min}} \int_{\eta_{\rm min}}^{\eta_{\rm max}} {d\eta\over\eta} \, S_a (\ln \eta)\ln(\eta/\eta_0)\,.
\end{equation}
In Fig.~\ref{plot:Gprime_filtered_WMAP7_20_vs_50PCs} we show an example with $\alpha=-0.026$ and compare 
the original linear $G'$ to the PC filtered $G'_{20}$. 
The filter introduces features at low and high $\eta$ that are not present in the actual source.
   Note that a Fisher analysis of sensitivity to $\alpha$ reveals that most of the signal to noise should lie in the $m_4$ component 
\cite{DvoHu10a} which carries the most significant deviations from zero in the data.

The $\chi^2$ analysis with all data implies $\alpha = -0.039\pm0.019$.  We can compare this result
to a direct MCMC analysis with $\alpha$ as a parameter constructed from 20 PCs: $\alpha =-0.027\pm0.021$.
Thus the simple $\chi^2$ approximation captures the information on $\alpha$ in the 20 PC posterior to 
$\sim 0.5\sigma$. 

\begin{figure}[tbp]
\includegraphics[width=0.45\textwidth]{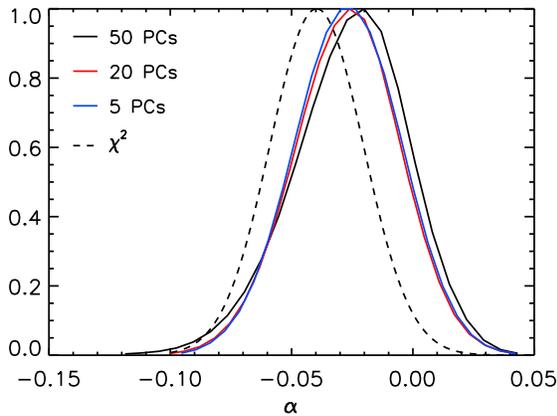}
\caption{Posterior probability distribution of $\alpha$  from a direct MCMC analysis constructed from $50$ PCs (black/solid curve), $20$ PCs (red curve), and $5$ PCs (blue curve). The distribution  from the $\chi^2$ approximation is shown in black/dashed curve.}
\label{plot:alpha_posteriors_alldata}
\end{figure}

We can further test the completeness of the 20 PC decomposition of $\alpha$ by going to 50 PCs.   In
this case $\alpha = -0.026\pm0.023$ showing that 20 PCs completely describe the observable properties
of $\alpha$.  
In fact, $5$ PCs are enough to describe the observable properties of $\alpha$ in this case; a direct MCMC analysis gives $\alpha=-0.026\pm 0.020$. Fig. \ref{plot:alpha_posteriors_alldata} shows that the full posterior distributions of $\alpha$ for these
cases are indistinguishable within the errors.  We also show the simple $\chi^2$ approximation which is shifted by $\sim 0.5\sigma$ as expected.

\begin{figure}[tbp]
\includegraphics[width=0.45\textwidth]{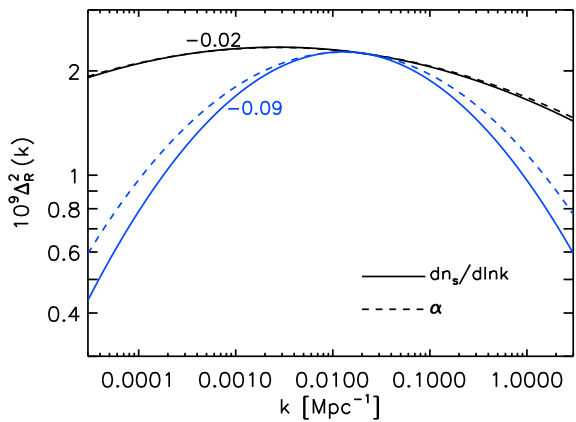}
\caption{Initial curvature power spectrum of a model with running of the tilt ($dn_s/d\ln k =$ $-0.02$, $-0.09$, solid curves) compared to a model with a linear deviation in $G'$ ($\alpha =-0.02$, $-0.09$, dashed curves).  For the $-0.02$ case, the two models are similar whereas for $-0.09$ the running of the tilt model has larger deviations from scale free conditions at low and high $k$.}
\label{plot:Pofk_running_Gprime_model}
\end{figure}

The posterior distributions are skewed to  negative values of $\alpha$.  
For example the ML model of the $50$ PC chain has  $\alpha=-0.021$ to be compared 
with a mean of $-0.026$.
 For large negative $\alpha$, the linear $G'$ model no longer matches
a running of the tilt due to the $I_1$ terms in 
Eq.~(\ref{eqn:alpharunning}).   In Fig.~\ref{plot:Pofk_running_Gprime_model}, we show an 
example with $n_0=0.96$ and $\eta_0 = 145$ Mpc for $\alpha= dn_s/d\ln k = -0.09$ and $-0.02$.  While the $\alpha$ model closely matches constant $dn_s/d\ln k$ 
for the smaller  value, it produces substantially less deviations at high and low $k$.
This bias explains the difference between constraints on the linear $\alpha$ model 
 and running of
the tilt found in \cite{DvoHu10a}.  For example, with the same data sets and priors
running of the tilt gives $dn_s/d\ln k=-0.018 \pm 0.019$.  Note the  ML $\alpha=-0.021$ 
from the 50PC chain is consistent with this constraint.

Another example is the step potential which has been employed to explain the
glitches in the power spectrum at $\ell \approx 20-40$
\begin{align}
V(\phi) = \frac{1}{2}m^{2}\phi^{2}\left[1+c\tanh\left(\frac{\phi - \phi_{s}}{d}\right)\right].
\label{eqn:step}
\end{align} 
For simplicity, we fix $\phi_s = 14.668$ so that the feature appears at the correct position 
to explain the glitches.   Although we set the smooth part of the potential to correspond to
an $m^2 \phi^2$ model with $m=7.126\times10^{-6}$ for the projection onto PCs, in the analysis we
retain the freedom to adjust the amplitude and tilt as usual.    This leaves us with 2 additional parameters
$c$ and $d$ to control the amplitude and width of the step.

\begin{figure}[tbp]
\includegraphics[width=0.45\textwidth]{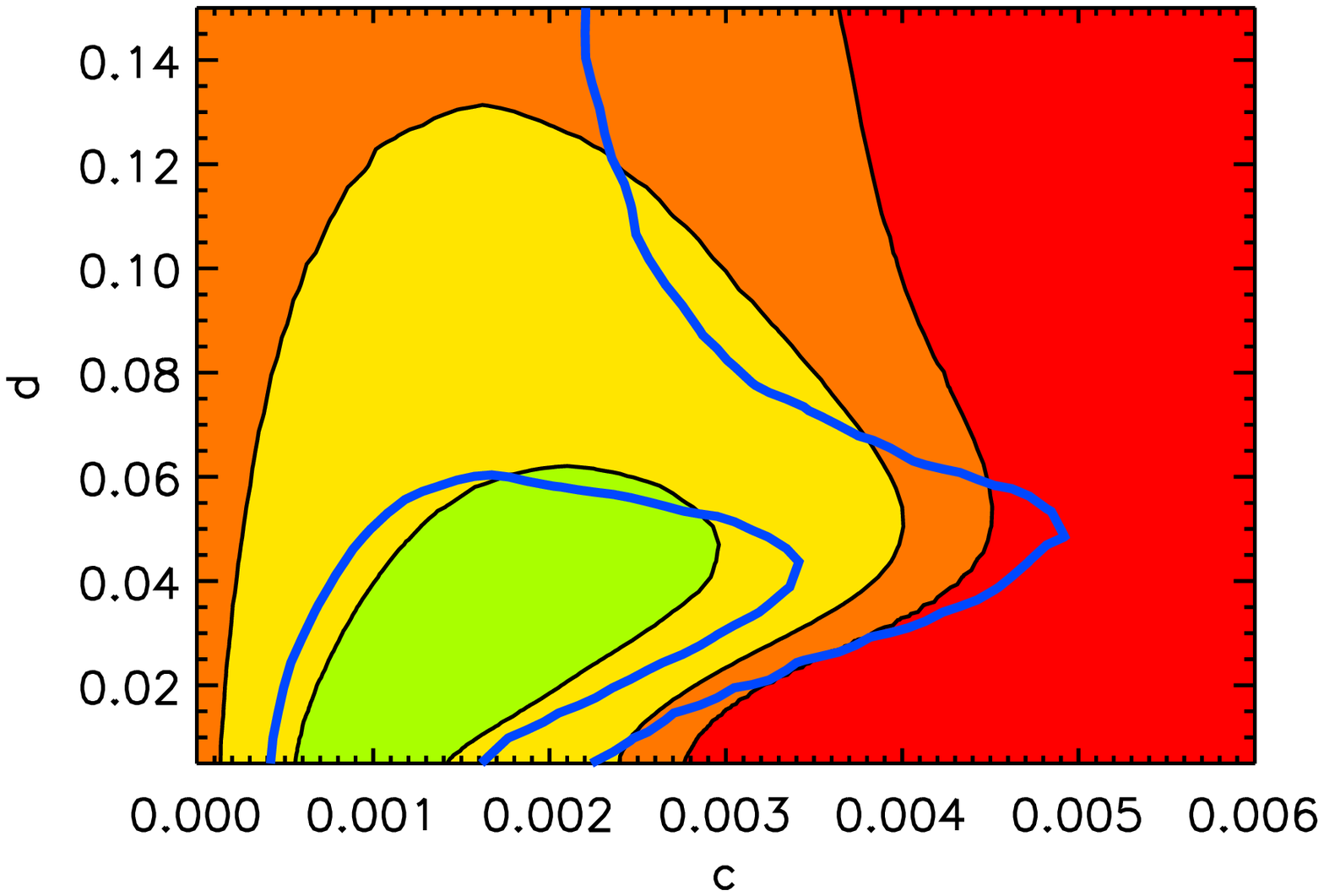}
\includegraphics[width=0.45\textwidth]{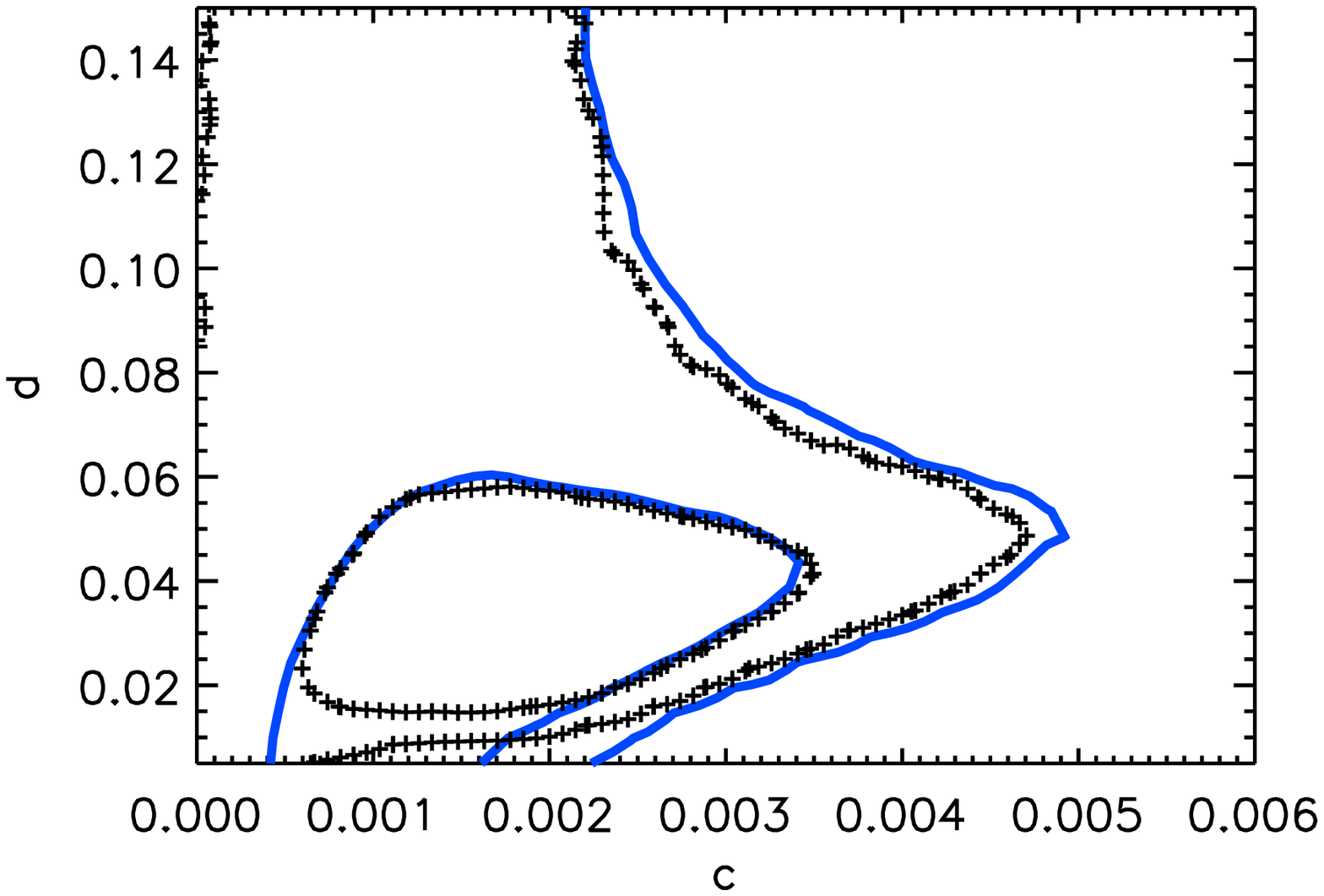}
\caption{Constraints on the step potential model parameters
$c$ (height of step) and $d$ width of step. {\rm Top panel}: the $\chi^2$ approximation (black curves) compared to the full $20$ PC posterior (blue curves). {\rm Bottom panel}: constraints from the $20$ PCs posterior (blue curves) compared to a direct GSR calculation of the model (black points).}
\label{plot:contour_final_range3}
\end{figure}

The constraints on $(c,d)$ from the $\chi^2$ approximation 
are shown in Fig.~\ref{plot:contour_final_range3} (top panel).    Note that the crude $\chi^2$ analysis correctly picks out the
favored parameters which can explain the glitches \cite{Mortonson:2009qv}.  The minimum $\chi^2$ model is $c=0.0015$, $d=0.026$ and is
favored over the PL $m_a=0$ (or $c=0$) model by $\Delta \chi^2 = -10.2$.  Although the
$\chi^2$ analysis assumes that the joint posterior in $m_a$ is a multivariate Gaussian, it does not make that assumption for parameter probabilities.  With the distorted shape of the confidence region, the $68\%$ contour corresponds to $\Delta \chi^2 = 2.5$, $95\%$ contour to $8.6$ and $99.7\%$ contour to $13.3$, compared with the more stringent $2.3$, $6.2$ and $11.6$ obtained for Gaussian distributions in $(c,d)$.   Here and below
we take a prior of $d>0.005$ due to our bandlimit of 1/4 efold (see below).

We again compare this with a full analysis of the joint 20 PC posteriors.   As in the case of $\alpha$, the  projection onto 
the two dimensional $m_a(c,d)$ space leaves us with too few samples in the original 20 PC chain
to reliably extract the posterior via importance sampling.  We instead run a direct MCMC analysis on the
20 PC description with $m_a(c,d)$.   These results are shown in Fig. \ref{plot:contour_final_range3} in blue points. 
The maximum likelihood model has $c=0.0016$, $d=0.025$ and is favored over PL $m_a=0$
(or $c=0$) by  $2\Delta \ln L= 9.1$.    These values are fully consistent with the simple
$\chi^2$ analysis.  
This improvement is a substantial fraction of the total of 17 available to the 20 PCs
from Tab.~\ref{table:alldata_20PCs_flat_priorI1} and is achieved with 3 parameters: $c,d$ and implicitly $\phi_s$, the location of the step.

\begin{figure}[btp]
\includegraphics[width=0.45\textwidth]{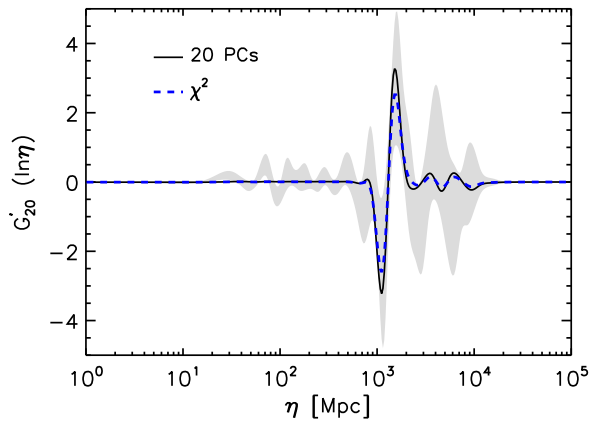}
\caption{The ML model of the step potential from the $\chi^2$ approximation is shown in blue dashed lines, and the ML model from the projection onto $20$ PCs [$m_a(c,d)$]  is shown in black lines. The step potential model captures the main feature seen in the fiducial all-data analysis (shaded 68\% CL area).}
\label{plot:Gprime_cdcontours_ML_20PCs_vs_Gaussian_analysis}
\end{figure}

The filtered $G_{20}'$ source for both the ML and minimum $\chi^2$ model
are shown in Fig.~\ref{plot:Gprime_cdcontours_ML_20PCs_vs_Gaussian_analysis} and
are consistent with the posteriors of the fiducial all data analysis.
Furthermore,  the $\chi^2$ analysis correctly picks out the
best fit region and qualitatively recovers its distorted shape.  The main difference
is that the confidence region is slightly underestimated.

\begin{figure}[btp]
\includegraphics[width=0.45\textwidth]{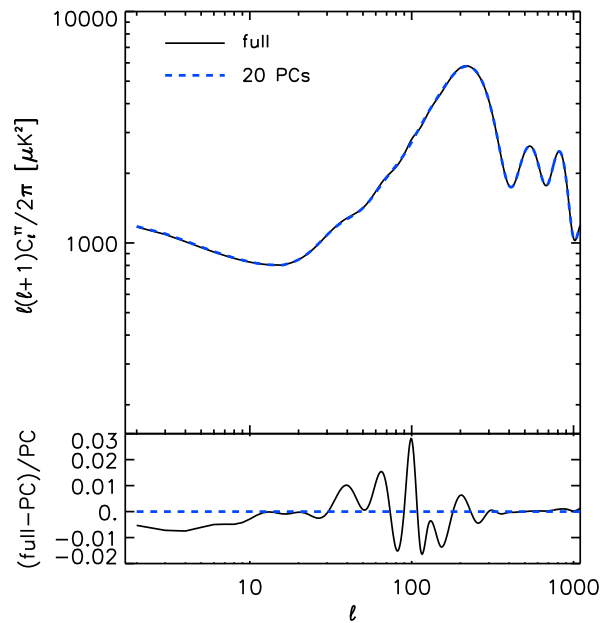}
\caption{{\rm Top panel:} step potential model with width $d=9.2\times 10^{-3}$ and height $c=4.6\times 10^{-4}$ represented by the full source function $G'$ (in black lines) compared to its $20$ PC description (in blue/dashed lines). {\rm Bottom panel}: fractional difference between the full GSR description and its $20$ PC decomposition. The oscillations at $\ell\sim100$ are not captured by the $20$ PCs. }
\label{plot:incompleteness_in_d}
\end{figure}

Finally, we test the completeness of the $20$ PC description of the step model by conducting
a separate MCMC with the full function $G'$ directly (see Appendix~\ref{sec:GSRaccuracy}, Eqs.~\ref{eqn:stepsource}-\ref{eqn:stepGprimebar} for details). The maximum likelihood model has $c=0.0021$, $d=0.029$ and is favored over PL by  $2\Delta \ln L=9.5$     As shown in Fig. \ref{plot:contour_final_range3} (bottom panel), the main difference
is that the models are more tightly constrained at $d < 0.01$.  The features in 
$G'$ span less than $\sim 1/4$ of an efold for these models and consequently the 20 PC 
decomposition is not complete.  
In Fig. \ref{plot:incompleteness_in_d} (top panel) we show a model with $d=9.2\times 10^{-3}$ and $c=4.6\times 10^{-4}$ represented by the full function $G'$ (in black lines) compared to its $20$ PCs description (in blue/dashed curves). The fractional difference between these two constructions is shown in the bottom panel.   In such models, the oscillations in the temperature power spectrum continue to higher $\ell$, in this case $\ell \sim 100$, and
are not allowed by the data.  

This example shows that the main limitation of the 20 PC analysis is
that it is too conservative for models with  high frequency structure in the source: such models tend to be in conflict with the data in ways not represented by the principal components.

\section{Discussion}
\label{sec:discussion}

We have conducted a complete study of constraints from the WMAP7 data on inflationary features beyond the slow roll limit.     Using a principal component (PC) basis that accommodates  order unity
features as fine as 1/10 of a decade  across more than 2 decades of the inflationary expansion, we find no significant deviations from slow roll.    Although one component
shows a deviation at the 98\% CL, it cannot be considered statistically significant given the
20 components tested.   
The maximum likelihood PC parameters only improves $2\Delta \ln L$ by 17 for the 20 
parameters added.

On the other hand, specific inflationary models may access this improvement with fewer
physical parameters.   Most of the improvement comes from fitting features in the temperature power
spectrum at multipoles $\ell \le 60$ with the known glitch at $20 \le \ell \le 40$ comprising
a large fraction.   

From our analysis, we also extract predictions for the corresponding features
in the polarization power spectrum that can be used to test their inflationary origin independently of a specific choice for the inflaton potential (cf. \cite{Mortonson:2009qv}).   In particular, one expects  a 
$\sim 26\%$  enhancement in the $EE$ power spectrum at $\ell = 39$ and a $\sim 37\%$ deficit around $\ell =25$ if the temperature features have an inflationary origin. 
Outside of the range of these low $\ell$ features, the predictions are very precise and any
violation of them in future observations would falsify single field inflation independently of
the potential.  

Our constraints can also be used to test any single field model that satisfies our conditions.  
Most of the information from the likelihood analysis is distilled in the means and covariance
of the principal components themselves which we make publicly available \footnote{\url{http://background.uchicago.edu/wmap_fast}}.   Two models illustrate this encapsulation: a linear source model that approximates running of the tilt and
a step potential model that fits the features at $\ell = 20-40$.   A simple $\chi^2$ analysis approximates the joint parameter posteriors despite its highly non-Gaussian form  for the step parameters.  
  This procedure greatly simplifies the testing of
inflationary models with features in that parameter constraints on any model that satisfies our conditions can be simply approximated without a case-by-case likelihood analysis.

\smallskip
{\it Acknowledgments:} We thank Peter Adshead and Mark Wyman for useful conversations.
This work was supported by the KICP under
NSF contract PHY-0114422.   WH was additionally supported by DOE contract DE-FG02-90ER-40560 and the Packard Foundation.
\appendix

\appendix

\section{MCMC Optimization}
\label{sec:MCMCopt} 
\subsection{Parameterization}

We seek to define amplitude and tilt parameters for the MCMC that are nearly orthogonal
to the PC parameters in order to improve the convergence properties of the MCMC chains.

A constant $G'$ is equivalent to tilt $n_s$ and hence PC components that
have long positive or negative definite stretches become degenerate with
tilt and cause problems for MCMC convergence.
Instead of a constant tilt, 
 we define a new chain parameter to be the
 average of $G'$ across a narrower range
 that is better associated with the observables
\begin{eqnarray}
\bar G' &=&
 {1 \over \ln\eta_{2}-\ln\eta_{1}}\int_{\eta_{1}}^{ \eta_2} {d \eta\over \eta} \, G' \,,
\label{eq:barGprime}
\end{eqnarray}
where specifically, we choose $\eta_1=30$ Mpc and $\eta_2=400$ Mpc to 
roughly minimize the variance of $\bar G'$ in the chain
 (see Fig.~\ref{plot:Gprime_posterior_WMAP7_BICEP_QUAD_20PCs_SN_H0_BBN_flat}).

Next, we replace the normalization parameter $G(\ln\eta_{\rm min})$ with 
\begin{equation}
A_s \equiv \ln \Delta_{\cal R}^2(k_p) \,,
\end{equation}
where in practice we choose $k_p=0.05$ Mpc$^{-1}$.

The effective tilt and normalization parameters bring the model of the power
spectrum from Eq.~(\ref{eqn:ourGSR}) to
\begin{eqnarray}\label{eq:final_powerspectrum}
\ln\Delta_{\cal R}^2 &=& \ln \left[ A_s \left( {k \over k_p} \right)^{-\bar G'}\right] +  \sum_{a=1}^{N}m_a[\bar W_a(k)-\bar W_a(k_p)] \nonumber\\
&& + \ln\left[ 1 + {1 \over 2} \left( {\pi \over 2} \bar G' + \sum_{a=1} ^N m_a \bar X_a(k)\right)^2\right] \nonumber\\
&&
- \ln\left[ 1 + {1 \over 2} \left( {\pi \over 2} \bar G' + \sum_{a=1} ^N m_a \bar X_a(k_p)\right)^2\right],
\end{eqnarray}
where
\begin{eqnarray}
\bar W_a(k) &=& \int_{\eta_{\rm min}}^{\eta_{\rm max}} {d\eta\over \eta}  W(k\eta) (S_a(\ln \eta) - \bar S_a) \,,\nonumber\\
\bar X_a(k) &=& \int_{\eta_{\rm min}}^{\eta_{\rm max}}  {d\eta\over \eta} X(k\eta) (S_a(\ln \eta) - \bar S_a) \,,
\end{eqnarray}
and
\begin{equation}
\bar S_a \equiv {1 \over \ln\eta_{2}-\ln\eta_{1}}\int_{\eta_{1}}^{ \eta_2} {d\eta \over \eta} S_a \,.
\end{equation}

Note that we can recover the tilt $n_s$, equivalent to the average of $\bar G$ across the whole range $\eta_{\rm min}$
to $\eta_{\rm max}$,  as 
\begin{equation}\label{eq:ns_derived}
n_s = (1-\bar G') + \sum_{a=1}^{20} m_a \bar S_a \,,
\end{equation}
and keep it as a derived parameter in the chain

Given the oscillatory nature of the $k$-space response to the PC eigenfunctions
through $\bar W_a$ and $\bar X_a$ and the geometric projection from $k$ to 
angular multipole $\ell$, normalization at a given $k$ does not correspond simply to
normalization at a given  $\ell$.   Since the observations best constrain the amplitude
of the temperature power spectrum near the first acoustic peak at $\ell \sim 220$
it is advantageous to use an $\ell$-space normalization in the MCMC and then transform
back to $A_s$.

Let us define a phenomenological parameter $A_c$ which renormalizes the
angular power spectra as
\begin{equation}\label{eq:cl_normalization}
C_\ell^{XY} = e^{\ln A_c}  {C_{220}^{TT\rm fid} \over \tilde C_{220}^{TT}}\tilde C_\ell^{XY}.
\end{equation}
Here $C_{220}^{TT\rm fid}$ is the temperature power spectrum at the first peak
of a fiducial model that fits the WMAP7 data. We use $C_{220}^{TT\rm fid}=0.747  \mu$K$^2$.
Thus if $A_c=0$, $C_{220}^{TT}= C_{220}^{TT\rm fid}$ regardless of the
PC parameters.

We can recover constraints on the $k$-space normalization by considering $A_s$ 
as a derived parameter.
If we compute the original 
 $\tilde C_\ell^{XY}$ with the $A_s=A_s^{\rm fid}$ of the fiducial model, then
 the true $A_s$
 is given by 
\begin{equation}
\ln A_s = \ln A_c + \ln( C_{220}^{TT\rm fid}/\tilde C_{220}^{TT}) + \ln A_s^{\rm fid}.
\end{equation} 
 In summary, we replace the parameters $n_s$ and $G(\ln\eta_{\rm min})$ with
 $\bar G'$ and $A_c$ in order to reduce parameter degeneracies that would otherwise inhibit chain convergence.

\subsection{Likelihood corrections}

To speed up the calculation of the WMAP and other CMB likelihoods
we employ three approximations when running the chains.  Firstly, we use a fitting function for the low
$\ell$ WMAP7 polarization likelihood as described in \cite{DvoHu10a}.  
Secondly, we calculate the CMB power spectra with gravitational lensing
artificially turned off.  Thirdly, we use the default $\ell$-space sampling of 
CAMB that is designed for smooth underlying power spectra.   
Each of these approximations produce small errors in the likelihood
evaluation that we can correct by importance sampling the chain.

The advantage of correcting these approximations in a postprocessing step is twofold.
The chains may be thinned due to the high correlation between samples in the
chain.  Secondly, postprocessing elements of the thinned chains
is embarrassingly parallel unlike the running of the original chain.

In practice, when we satisfy our convergence criterion described in the main text, we thin the chains by a factor of half of the correlation length. 
We have tested that with such thinning we reproduce the posteriors of the original chains. 
Next 
we compute the CMB power spectra of the thinned chains with lensing turned on and a higher $\ell$-space sampling (CAMB ``accuracy boost" $2$). We use these high accuracy power spectra to correct the chain multiplicity for the change in the likelihood.

\begin{figure}[btp]
\includegraphics[width=0.45\textwidth]{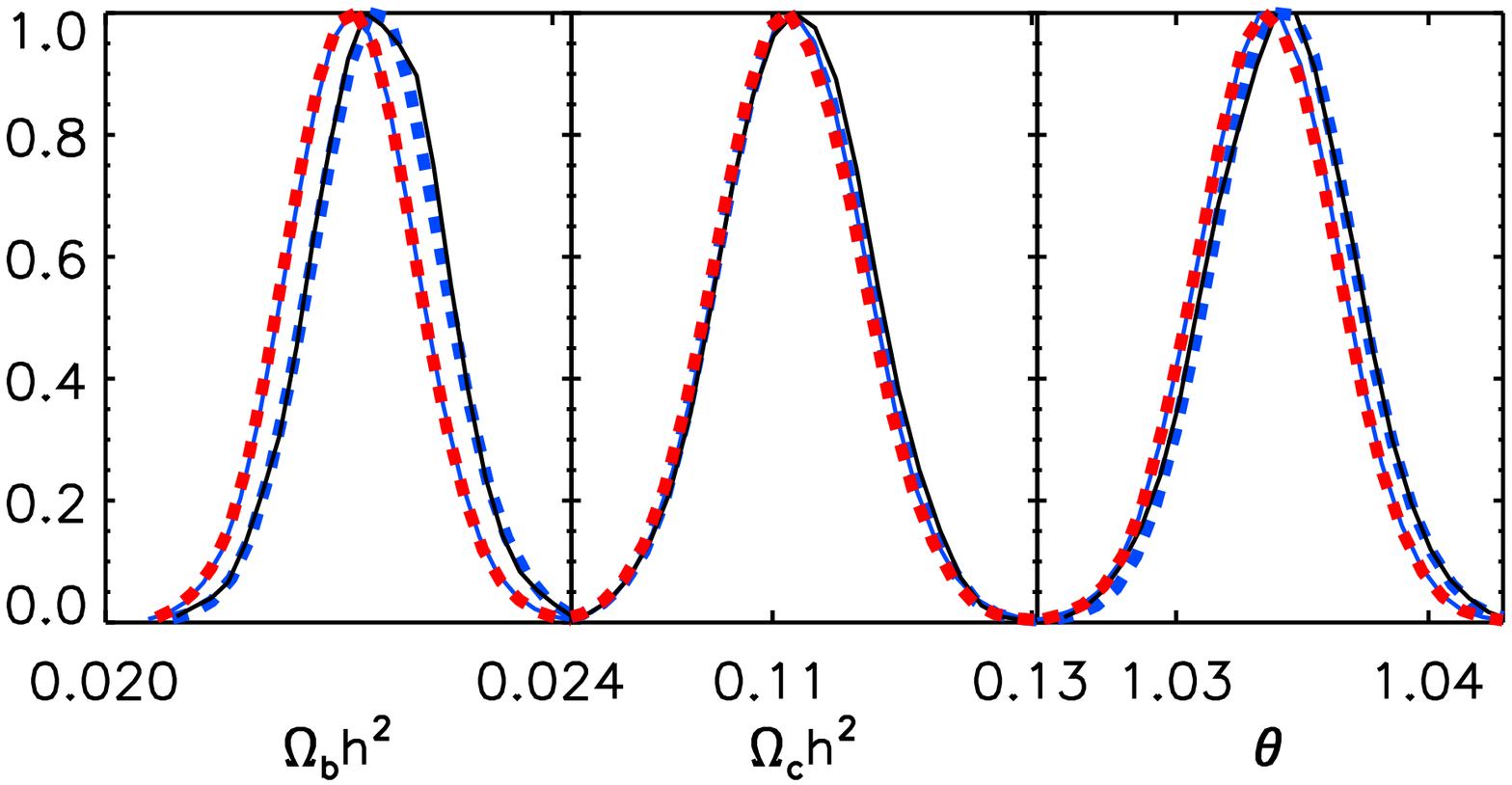}
\includegraphics[width=0.45\textwidth]{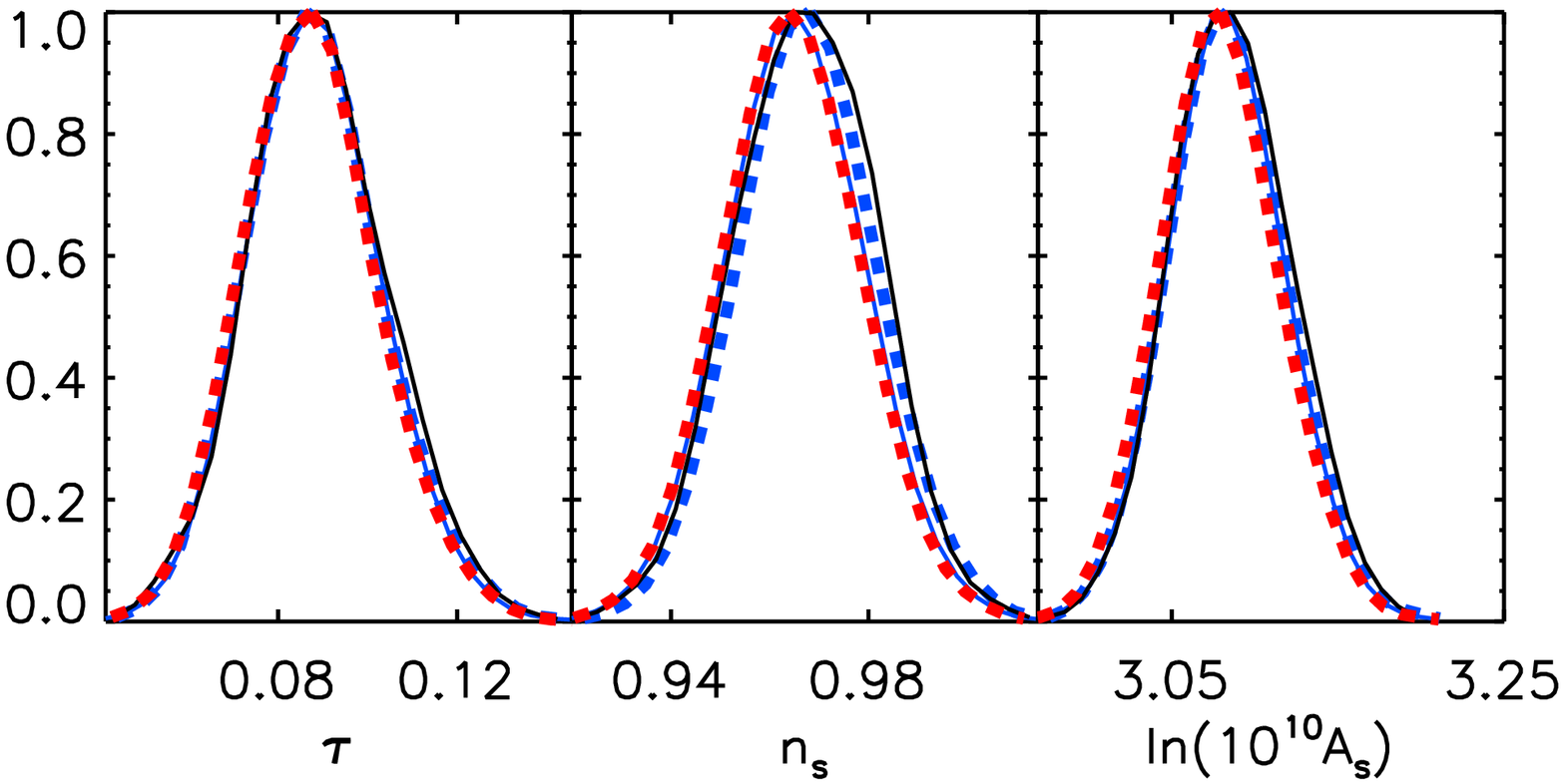}
\caption{Power law parameter posteriors  from the approximations used to run the MCMC chain (in blue/solid curve), from an independent MCMC with no approximation (in black/solid curve),  from the approximate chain with importance sampling correction
(in blue/dashed curve), and from the approximate chain without lensing
correction (in red/dashed curve).}
\label{plot:LCDM_postprocessing}
\end{figure}

In Fig.~\ref{plot:LCDM_postprocessing} we show as an example the posteriors coming directly from power law (PL) chains (in blue/solid curves), the chains with all corrections (in blue/dashed lines) and finally all corrections but lensing  (in red/dashed lines).  These should be compared with
results  from a separate chain run with all the corrections turned on from the start (in black/solid lines).   Importance sampling 
accurately models the impact of the small corrections for all parameters.   The leading correction is on $\Omega_b h^2$ from lensing.

\begin{figure}[btp]
\includegraphics[width=0.45\textwidth]{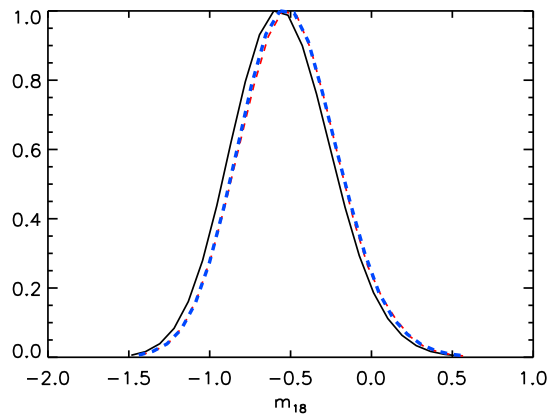}
\caption{\footnotesize The $m_{18}$ probability distributions  from the approximations used to run the MCMC with all data and $I_{1,{\rm max}}=1/\sqrt{2}$ (in black lines), from the approximate chain with importance sampling correction (in blue/dashed lines), and from the approximate chain without lensing correction (in red/dashed lines).  $m_{18}$ has the largest correction of the PC amplitudes which is still $\ll 1\sigma$ and dominated by the
 lensing correction.}
\label{plot:postprocessing}
\end{figure}

In Fig.~\ref{plot:postprocessing}, we show the impact of the corrections on the PC chain using $m_{18}$ as an example
with the largest correction.  The correction on PC parameters is extremely small and again dominated by lensing.

\section{GSR Accuracy}
\label{sec:GSRaccuracy}

To test the accuracy of the GSR approximation in the PC space, we need to consider the inverse problem: construct an inflationary model
that matches a desired $G'$ for which we can solve exactly for the curvature power spectrum.

In the forward direction, given an inflationary model  we can compute the exact curvature spectrum by first
evaluating the background behavior of the model through 
\begin{equation}
g(\ln\eta) = {f'' \over f} - 3{f' \over f} = {3 \over 2} G' + \left( {f' \over f} \right)^2\,,
\label{eq:littleg}
\end{equation}
and then solving the equation
\begin{equation}
{d^2 y \over dx^2} + \left( 1 - {2 \over x^2} \right) y = {g \over x^2} y \,,
\label{eqn:yeqn}
\end{equation}
where $x = k\eta$, subject to the usual  Bunch-Davies initial conditions.  The curvature power spectrum
is then given by 
\begin{equation}
\Delta_{\cal R}^2 = \lim_{x \rightarrow 0}  x^2 {|y|^2 \over f^2} \,.
\label{eqn:fullsolution}
\end{equation}

\begin{figure}[htbp]
\includegraphics[width=0.45\textwidth]{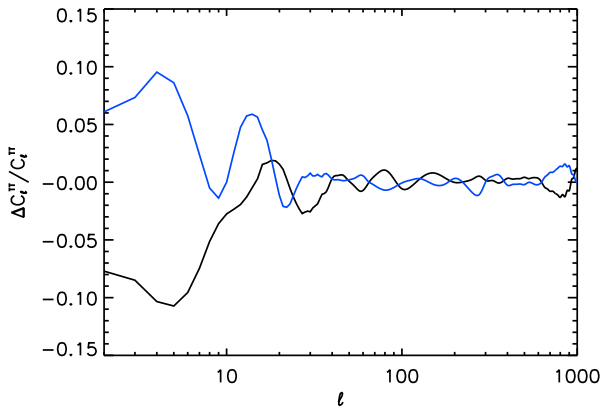}
\caption{Fractional difference in temperature power spectra between GSR and the exact inflationary solution for the maximum likelihood model from the all-data analysis (in black lines) as well as a model that saturates the prior $I_{1,\rm max}=1/\sqrt{2}$ from the chain (in blue lines).
For reference, the ML model has $I_{1, \rm max}=0.66$.}
\label{plot:DeltaClTT_ML_LR_SR}
\end{figure}

\begin{figure}[htbp]
\includegraphics[width=0.45\textwidth]{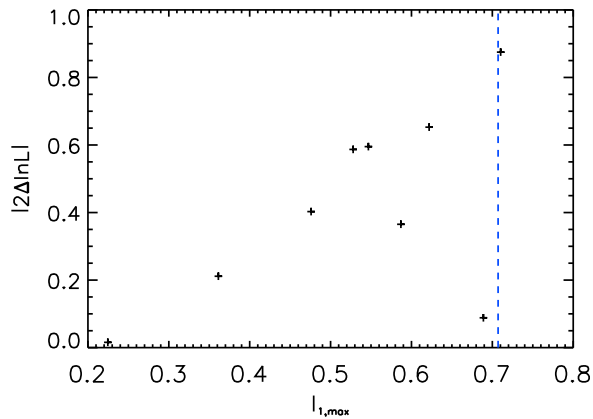}
\caption{Likelihood difference between the GSR solution and the full inflationary calculation of a series of step potential models as a function of $I_{1,\rm max}$.  Models were chosen from the full GSR chain to be the maximum likelihood in a series of bins in step amplitude $c$.  The maximal error is small below  $I_{1,\rm max}=1/\sqrt{2}$ (blue dashed line), the prior in the fiducial all-sky analysis.}
\label{plot:I1max_vs_Deltachi2}
\end{figure}

Therefore to test the GSR approximation we first need to determine the function $g$
that matches a given $G'(\ln\eta)$.
Transforming variables to $r=f'/f$, we obtain  from Eq.~(\ref{eq:littleg})
\begin{equation}
r' - 3r = {3 \over 2} G' \,,
\end{equation}
which has the general solution
\begin{equation}
r = {3 \over 2}\eta^3 \int {d \tilde \eta \over \tilde\eta}  \tilde \eta^{-3} G' + C \eta^3 \,.
\end{equation}
Let us choose the integration constant 
\begin{equation}
C = -  {3 \over 2} \int_{\eta_{\rm min}}^{\eta_{\rm max}}{d \tilde \eta \over \tilde\eta} \tilde \eta^{-3} G'  \,,
 \end{equation}
 and assume $G'$ vanishes outside this range.  We then get
 \begin{equation}
 r = -{3 \over 2} \eta^3 \int_{\eta}^{ \eta_{\rm max}} {d \tilde \eta \over \tilde\eta} \tilde \eta^{-3} G' \,,
\end{equation}
for $\eta > \eta_{\rm min}$ and
\begin{equation}
r =  -{3 \over 2} \eta^3 \int_{\eta_{\rm min}}^{\eta_{\rm max}}{d \tilde \eta \over \tilde\eta}\tilde \eta^{-3} G' \,,
\end{equation}
for $\eta < \eta_{\rm min}$. 
 With this numerical solution we construct $g$ as
 \begin{equation}
 g = {3 \over 2} G' + r^2 \,.
 \end{equation}
 This suffices to specify the source for $y$ in Eq.~(\ref{eqn:yeqn}).  
  Finally, to get the curvature power spectrum we need $f$ at some 
 $\eta_{\rm lim} \ll k_{\rm max}^{-1}$.   However since this quantity is independent of
 $k$, it is absorbed into our normalization definition.
In Fig.~\ref{plot:DeltaClTT_ML_LR_SR}, we take parameters from the all-data chain and use this
technique to calculate the temperature power spectra of matching inflationary models exactly.  
Even for the model that saturates  the $\Imax=1/\sqrt{2}$ prior, the WMAP likelihood difference between
the exact and GSR calculation is  $|2\Delta \ln L|=0.4$.

Using the step model chain from \S \ref{sec:model}, we can explore the accuracy of the
GSR approximation as a function of $\Imax$ independently of the prior taken in the all data analysis.
Specifically, we take a model from the chain that defines $G'$
and construct the matching full inflationary model as above.

Recall that to
construct $G'$, we solve for the background evolution of $\phi$ in the  step potential of Eq.~(\ref{eqn:step}). This specifies the $m^2 \phi^2$ model source through 
Eq.~(\ref{eqn:sourcefunction}), which we call $G'_m$.   To allow
for a retilting of the spectrum, we add an extra constant parameter $\bar G'_p$ to the model source
to form the full source
\begin{equation}
G'(\ln\eta; c,d,\bar G'_p)=G'_m(\ln\eta;c,d) + \bar G'_p.
\label{eqn:stepsource}
\end{equation}
The GSR approximation then tells us that the curvature spectrum is given by 
\begin{eqnarray}
\ln\Delta_{\cal R}^2  =  \ln \left[ A_s \left( {k \over k_p} \right)^{-\bar G'_p}\right]
+ I_m(k)-I_m(k_p)
\end{eqnarray}
with
\begin{eqnarray}
 I_m(k) &=&  \int_{ \eta_{\rm min}}^{\eta_{\rm max}}{d \eta \over \eta}
W(k\eta) G'_m  \\
&&  + \ln \Bigg[ 1+ {1\over 2} \left({\pi \over 2} \bar G'_p + \int_{\eta_{\rm min}}^{\eta_{\rm max}} {d\eta \over \eta} X(k\eta) G'_m\right)^2 \Bigg]. \nonumber
\end{eqnarray}
In Fig.~\ref{plot:I1max_vs_Deltachi2}, we compare the impact of taking this power spectrum
to a full inflationary calculation with matching source (\ref{eqn:stepsource}) on the
WMAP likelihood.   For the full calculation of Eq.~(\ref{eqn:fullsolution}), we take 
$\Delta_{\cal R}^2(k_p)=A_s$ to define the normalization $f$.
Since
$\Imax$ increases monotonically with $c$, we show models with maximum likelihood parameters in uniform bins of $c$.   Note that the maximal error increases with $\Imax$ but does not exceed order unity at $\Imax< 1/\sqrt{2}$.

For reference, to compute a matching 20 PC representation as in Fig.~\ref{plot:incompleteness_in_d} we take the  amplitudes of the principal components from Eq.~(\ref{eq:mas}) and use Eq.~(\ref{eq:barGprime}) to define 
\begin{equation}
\bar G' = \bar G'_p +
 {1 \over \ln\eta_{2}-\ln\eta_{1}}\int_{\eta_{1}}^{\eta_2} {d\eta\over\eta} \,  G'_m(\ln \eta)\,.
 \label{eqn:stepGprimebar}
\end{equation}
and keep the normalization $A_s$ fixed.

 \vfill
 
\bibliographystyle{arxiv_physrev}
\bibliography{PCfull}

\end{document}